\documentclass[journal]{IEEEtran}
\usepackage{amssymb,epsfig,color,cite,amsmath,subfigure,amsfonts,fancybox,mathrsfs}
\usepackage{balance}

\usepackage{array}
\newcolumntype{P}[1]{>{\centering\arraybackslash}p{#1}}

\usepackage{float}

\newcommand{\diag}{{\rm diag}}
\newcommand{\wdt}{\widetilde}
\newcommand{\ol}{\overline}
\newcommand{\wdh}{\widehat}

\newcommand{\bed}{\begin{displaymath}}
\newcommand{\eed}{\end{displaymath}}
\newcommand{\bea}{\bed\begin{array}{rl}}
\newcommand{\eea}{\end{array}\eed}
\newcommand{\beq}[1]{\begin{equation} \label{#1}}
\newcommand{\eeq}{\end{equation}}
\newcommand{\barray}{\begin{array}{ll}}
\newcommand{\earray}{\end{array}}
\newcommand{\disp}{\displaystyle}
\newcommand{\ad}{&\!\!\!\disp}

\newcommand{\e}{{\epsilon}}

\def\clC{{\cal C}}

\def\clE{{\cal E}}

\def\clN{{\cal N}}

\def\clS{{\cal S}}

\def\clV{{\cal V}}

\def\e{{\epsilon}}

\def\Base{{\rm Base}}
\def\Tr{{\rm Trace}}
\def\Range{{\rm Range}}

\def\Rank{{\rm Rank}}

\newcommand{\dbR}{{\Bbb R}}
\newcommand{\dbC}{{\Bbb C}}

\newcommand{\dbU}{{\Bbb U}}

\def\one{{\hbox{1{\kern -0.35em}1}}}

\newtheorem{thm}{Theorem}

\newtheorem{lem}{Lemma}
\newtheorem{rem}{Remark}

\newtheorem{exm}{Example}
\newtheorem{asm}{Assumption}

\begin{document}
\title{Stochastic Hybrid System Modeling and State Estimation of Modern Power Systems under Contingency}

\author{Shuo Yuan,\thanks{Shuo Yuan is with the Department of Electrical and Computer Engineering,
	Wayne State University, Detroit, Michigan 48202, USA {\tt\small <shuoyuan@wayne.edu>}}
Le Yi Wang,\thanks{Le Yi Wang is with the Department of Electrical and Computer Engineering,
	Wayne State University, Detroit, Michigan 48202, USA {\tt\small <lywang@wayne.edu>}} \IEEEmembership{Life Fellow, IEEE,}
George Yin,\thanks{George Yin is with the Department of Mathematics,
University of Connecticut,
Storrs, Connecticut 06269-1009, USA {\tt\small
<gyin@uconn.edu>}} \IEEEmembership{Life Fellow, IEEE,}
Masoud H. Nazari,\thanks{Masoud Nazari is with the Department of Electrical and Computer Engineering,
Wayne State University, Detroit, Michigan 48202, USA {\tt\small <masoud.nazari@wayne.edu>}}
\IEEEmembership{Senior Member, IEEE}

\thanks{}}
\date{}
\maketitle

\begin{abstract}
This paper introduces a stochastic hybrid system (SHS) framework in state space model to capture sensor, communication, and system contingencies in modern power systems (MPS). Within this new framework, the paper concentrates on the development of state estimation methods and algorithms to provide reliable state estimation under randomly intermittent and noisy sensor data. MPSs employ diversified measurement devices for monitoring system operations that are subject to random measurement errors and rely on communication networks to transmit data whose channels  encounter random packet loss and interruptions. The contingency and noise form two distinct and interacting stochastic processes that have a significant impact on state estimation accuracy and reliability. This paper formulates stochastic hybrid system models for MPSs, introduces coordinated observer design algorithms for state estimation, and establishes their convergence and reliability properties.
 A further study reveals a fundamental design tradeoff between convergence rates and steady-state error variances. Simulation studies on the IEEE 5-bus system and IEEE 33-bus system are used to illustrate the modeling methods, observer design algorithms, convergence properties, performance evaluations, and impact sensor system selections.
\end{abstract}


\section{Introduction}

\IEEEPARstart{M}{odern} power systems (MPS) employ advanced technologies in sensing, communications, control, and management \cite{KN,GS}. In particular, in microgrids with high penetration of renewable generations and controllable loads, sophisticated compensation and control systems have been employed \cite{Z,NIL,H}. To facilitate such technological advances, upgraded sensing systems such as phasor measurement units (PMUs) and wireless communication systems, including local, regional, and satellite communication networks, have become commonplace and have greatly expanded the capability of data acquisition, exchange, and processing \cite{PMU1,PMU2}. The new measurement capability enables energy management systems (EMS) to treat power systems as dynamic systems and to estimate internal states for control and management.

Power system management relies on signals and data on generators, users, buses, and transmission systems for optimization,  stabilization, and optimal control \cite{Yi2016Initialization, Dall2013Distributed,Nazari2014Distributed,Nazari2020Communication,
Angelo, Guttromson, Donnelly, Cardell1, Dou2022, Kasis2022, Sek2023}. These methodologies and algorithms assume the availability or direct measurements of all variables including state variables and others. However, measuring all state variables in  microgrids (MGs) and on renewable generators is costly. In addition, signal measurements and communication systems always encounter stochastic uncertainties, including errors and interruptions. In particular, wireless communications inherently involve random uncertainties such as channel interruption, packet loss, and delays, which may create random link interruptions and consequently randomly switching communication networks \cite{lapidoth1994performance}. Capturing random communication uncertainties leads to time-varying network topology  \cite{moreau2005stability}, coordinated control/communication co-design \cite{XWYZ}, delay models in block erasure channels \cite{Delay3},  random interruption models \cite{Tru2022}. In our earlier work \cite{Nazari2021Impact, Xie2021Impact}, we investigated the impact of communication packet loss and noisy environment on the performance of optimal load tracking and allocation in DC MGs.

This paper introduces a general stochastic hybrid system framework in state space model to capture sensor, communication, and system contingencies in MPS, and studies in particular the impact of sensing reliability on the capability, error, and reliability of internal state estimation in MPS. Sensors are subject to measurement errors and wireless communication systems have reliability limitations that are characterized by random packet losses. These stochastic uncertainties have a critical impact on state estimation accuracy and reliability. This paper models measurement errors as Brownian motions, and system interruptions and contingencies as discrete events, leading to stochastic hybrid system (SHS) models.
 The framework converts a physical power grid into a virtual, dynamic, state-space model that
provides opportunities to estimate internal (unmeasured) states by using only a few sensors. System dynamics and time trajectories from these sensors provide rich information for state estimation. Under sensing contingency and data interruption, the algorithms of this paper integrate estimation from different observers to derive a convergent estimator for the entire state.

State estimation problems were investigated in power systems, mostly without stochastic switching uncertainties. For example, extended Kalman filter (EKF) algorithms were employed in \cite{Teb2015}  for state estimation in power systems using PMU data.  Control and protection applications of dynamic state estimation were studied in \cite{Liu2021}.  A decentralized algorithm for real-time estimation of dynamic states of power systems was proposed in \cite{Sin2014}. Dynamic models were derived in \cite{Loo2013} on synchronous and induction generators for wind turbines using state-space representations.
The methods and algorithms of this paper do not introduce additional limitations of scalability of state estimation problems, but rather enhance state estimation algorithms and their reliability under stochastic uncertainties. In this framework, accuracy and reliability of state estimation are critically affected by stochastic processes and become a critical issue for reliable power system management.

Mathematically, randomly switched systems can be modeled and treated as stochastic hybrid systems \cite{Seah2009}, stochastic systems with time-varying parameters \cite{Dragan}, or hybrid switching diffusions \cite{LP, Cassandras, YinZ10, Lunze, Teel5}. Here we investigate the possibility of using only limited sensors to achieve the goals of state estimation, introduce coordinated observer design methods to overcome sensor interruptions, establish convergence properties under random noise and random switching, and evaluate inherent design tradeoffs.
In our previous theoretical work \cite{WY1, WY2, WY3}, we introduced new methodologies for designing convergent subsystem observers on their observable sub-states in randomly switched linear systems (RSLSs), and achieved global convergence results. This paper aims to investigate potential applications of these theoretical foundations on state estimation problems in MPS and study further reliability issues under randomly intermittent and noisy sensor data.

The main contributions of this paper are summarized as follows:
\begin{enumerate}
	\item This paper proposes a new SHS framework to model modern hybrid systems under contingencies. From physical power system networks that involve both dynamic and non-dynamic buses,  virtual linearized dynamic models of MPS are developed that contain only dynamic nodes in state space models. The SHS model accommodates the practical and common scenarios in MPS in which contingency  often  results in unobservable systems.
	
	\item This paper investigates a coordinated observer design method and related algorithms that can accommodate unobservable subsystems due to contingency and achieve convergence for the entire state.
	
	\item This paper establishes bounds on the sampling and data processing time interval so that by using a time interval below the bound, convergent observers can be designed. Convergence rates and steady-state estimation error variances are derived.

\item This paper  highlights an inherent design tradeoff on observer feedback gain selection that must balance convergence rates and steady-state error variances.
	
	\item This paper uses the IEEE 5-bus system and IEEE 33-bus system to illustrate different model derivation approaches, validate RSLS models, observer design, convergence properties, design tradeoffs, sensor selections, and relationships between sensor reliability and state estimation accuracy and reliability.
\end{enumerate}

The rest of the paper is organized as follows. Section \ref{P} defines the notation and formulates the state estimation problems in dynamic power systems. Section \ref{SSM} introduces the SHS modeling approach for deriving virtual dynamic state-space models for MPS.  Sensor systems, contingencies, and their SHS models are described in Section \ref{Sensors}. Section \ref{Design} discusses observer design procedures and algorithms, and establishes main convergence analysis and estimation error calculations. Performance evaluation case studies are presented in Section \ref{case} by using two IEEE bus systems, illustrating different modeling methods, scalability, and performance of the framework, methods, and algorithms. Different sensor systems, their reliability, and their impact on state estimation errors are detailed in this section. The main conclusions of this paper are summarized in Section \ref{con}.

\section{Preliminaries}\label{P}

Denote by
$\dbR$
the field of real numbers and $\dbC$
the field of complex numbers. For a column vector $v\in \dbR^n$, $\|v\|$ is its Euclidean norm. For a matrix $M\in \dbR^{n\times m}$, $M^{\top}$ is its transpose,  $\Tr(M)$ is its trace, $\lambda(M)$ is an eigenvalue of $M$,  $\sigma(M)=\sqrt{\lambda(M^{\top}M)}$ is a singular value of $M$,
$\sigma_{\min}(M)$ is its minimum singular value,  and $\sigma_{\max}(M)$ is its largest singular value. Its operator norm induced by the Euclidean norm is
$\sigma_{\max}(M)=\|M\|=\sup_{\|v\|=1} \|Mv\|.$
The  kernel or null space of $M\in \dbR^{n\times m}$ is
$\ker(M)=\{x\in \dbR^m: Mx=0\}$ and its range is $\Range(M)=\{y=Mx: x\in \dbR^m\}$.
For a subspace $\dbU\subseteq \dbR^n$ of dimension $p$, a matrix $M\in \dbR^{n\times p}$ is said to be a base matrix of $\dbU$, written as $M=\Base(\dbU)$,  if the column vectors of $M$ are linearly independent and ${\rm Range}(M)=\dbU$.
Denote $\diag[M_1,\ldots, M_\rho]$ as the block diagonal matrix of (not necessarily square) matrices $M_i,i=1,2,\cdots,\rho$.

A function $y(t)\in \dbR$ in a time interval $[0,\tau)$ is piecewise continuously differentiable if $[0,\tau)$ can be divided into a finite number of subintervals $[t_{k-1},t_k)$, $k=1,\ldots, l$, $t_0=0$, $t_l =\tau$ such that $y(t)$ is right continuous in  $[t_{k-1}, t_k)$ and continuously differentiable, to any order as needed, in  $(t_{k-1}, t_k)$. The space of such functions is denoted by $\clC[0,\tau)$.

This paper will introduce an RSLS framework that can  represent dynamics of MPS under diversified contingencies. To be concrete and focused, this paper will use AC power grids in methodology and algorithm development and treat sensor and communication interruptions as representative contingencies in examples and case studies.

Voltages and currents in AC power systems will be represented by their phasors $\vec V=V \angle \delta$ and $\vec I=I \angle \gamma$. This paper studies state estimation by using various sensors including PMUs and frequency measurements,  and involving communication systems in data transmissions. Since PMU measurements are computed from noise-corrupted voltage and current waveforms and use wireless and satellite communications for obtaining universal time, they are inherently subject to random measurement errors and communication uncertainties such as random packet losses that are commonly characterized by probabilistic packet delivery ratios.

The physical power system under study is a networked system that contains $\gamma$ buses and whose physical transmission lines form naturally a physical network graph $\clN=\{\clV, \clE\}$ where $\clV$ is the set of buses (nodes in a graph) and $\clE$ is the set of feeder/transmission links (edges in a graph). The transmission line $(i,j)\in \clE$ is bi-directional, i.e., $(i,j)\in \clE \rightarrow (j,i)\in \clE$.
For Bus $i$, its neighbor $\clN_i$ is the set of buses $j$ that are connected to it, namely, $\clN_i = \{j\in \clV: (i,j)\in \clE \hbox{ or } (j,i)\in \clE \}$.

The main SHS framework and methodologies of this paper are general and can be used to model any bus systems in MPS under diversified contingency scenarios.  For concreteness, this paper is focused on sensor and communication interruptions as representative contingencies. Also, real power management  problems are used to explain mathematical expressions.

\section{Dynamic State Space Models of Microgrids}\label{SSM}

In our SHS framework, we classify buses in a power grid into two types: dynamic buses and non-dynamic buses, regardless of their physical configurations that may be generators, loads, energy storage systems, microgrids, or a cluster of combined facilities and users. This classification is different from the traditional power system analysis that classifies buses into three types: (1) PV buses that are typically for generators. (2) PQ buses that are commonly used to represent loads. (3) Slack buses that have reference voltage and angle with infinite real and reactive power to achieve power balance instantaneously. usually, generators have dynamics, such as swing equations,  and loads are nonlinear functions without dynamics.\footnote{In power systems, a bus with controllable power is called dispatchable. Dispatchable sources of electricity refer to electrical power sources that can adjust their power outputs on demand. In contrast, many renewable energy sources are intermittent and non-dispatchable. Also, if a load is controllable, the load power consumption  that can be controlled becomes dispatchable. In our SHS models, this classification affects only the input variables as control (dispatchable power) or disturbance (non-dispatchable power). It is independent of a bus to be dynamic or non-dynamic.}

However, in MPS with high penetration of renewable generators, controllable loads, and added local control systems and management strategies,  buses have more complicated characterizations. For example, a bus with controllable loads that involve a local PI controller to track electricity demands becomes a dynamic bus. For state estimation in SHS, the power system models must be put into a state space model whose state variables involve only dynamic buses. All other intermediate variables from non-dynamic buses must be represented as functions of state variables and inputs. Consequently, the final SHS model will be a virtual power grid that has only dynamic buses with combined links. Detailed derivations
for virtual SHS models of MPS will be presented next.

\subsection{General Dynamic Modeling Methods}

In this section, we first outline a general approach to convert a physical power grid involving both dynamic and non-dynamic buses into a {\it virtual dynamic grid} in which all buses are dynamic. The virtual dynamic power grid can then be represented by a state space model. In our SHS framework, for physical power grids all buses are classified into dynamic and non-dynamic buses.

\textbf{1) Dynamic Buses}

In general, if Bus $i$ is dynamic, then it entails a local state space model,
\beq{nd0}  \dot z^d_i = f_i(z^d_i, z_i^-, v_i^d, \ell^d_i),\eeq
where $z^d_i$ is the local state variable, $z_i^-$ is the neighboring variables of Bus $i$ which may be state variables of its neighboring dynamic buses, or intermediate variables of its neighboring non-dynamic buses, $v_i^d$ is the local control input, and $\ell^d_i$ is the local load. Note that  the work ``load" is used to represent power outputs/consumptions of the assets that cannot be actively controlled, such as regular loads, fixed-blade wind generators, solar panels, constant-charging-current batteries, etc. Similarly, the word ``control input" is the consolidated power of the assets whose power outputs or consumptions
 can be actively controlled, such as traditional generators, controllable loads, actively managed battery systems, tunable wind turbines, etc. Both the control input and load can be measured. Consequently, all dispatchable assets on a bus are grouped into $v_{i}^{d}$ and all non-dispatchable assets are grouped into $\ell^d_i$. If a bus does not have any dispatchable assets, then $v_i^d=0$ and the bus becomes totally non-dispatchable.

\textbf{2) Non-dynamic Buses}

If the $j$th bus is non-dynamic, then it is represented by an implicit algebraic relationship,
\beq{nd1} 0=g_j(z^{nd}_j,z_j^-, v_j^{nd},\ell^{nd}_j),\eeq
where $z^{nd}_j$, by an abuse of terminology,  is the local ``state variable" vector, $z_j^-$ is the neighboring variables, $v_j^{nd}$ is the local control input, and $\ell^{nd}_j$ is the local load. Note that in modern power systems many non-dynamic systems, such as smart lighting systems, smart appliance, have controllable power that is collectively represented by $v_{j}^{nd}$. In this scenario,  this non-dynamic bus is dispatchable. Otherwise, it is non-dispatchable. It should be cautioned that if one includes sensor dynamics, actuator dynamic systems  and controller dynamics (such as PID controllers) in a bus model whose physical part is non-dynamic, then the bus model will involve differential equations. As a result, the bus must be moved to the class of ``dynamic buses".

\textbf{3) Virtual Dynamic State Space Models}

Suppose that the $\gamma$ buses in $\clN$ contain $\gamma^d$ dynamic buses and $\gamma^{nd}=\gamma-\gamma^d$ non-dynamic buses.\footnote{Since this paper deals with state estimation under state space models, we assume that $1\leq \gamma^d\leq \gamma$, namely at least one bus is dynamics. But $\gamma^{nd}=0$ is possible, meaning that all buses are dynamic.} Without loss of generality, let the first $\gamma^d$ buses be dynamic.
Define the states, inputs, loads, and outputs from all buses,
\bea
z^d\ad =  \left[\begin{array}{c} z^d_1 \\ \vdots \\ z^d_{\gamma^d} \end{array} \right]~\hbox{State variables of dynamic buses},\\
z^{nd}\ad =  \left[\begin{array}{c} z^{nd}_{\gamma^d+1} \\ \vdots \\ z^{nd}_{\gamma} \end{array} \right] ~\hbox{State variables of non-dynamic buses},
\eea
\bea
z\ad =  \left[\begin{array}{c} z^d \\  z^{nd} \end{array} \right] ~\hbox{All variables},\vspace{0.3ex}\\
v^d\ad =  \left[\begin{array}{c} v_1^d \\ \vdots \\ v_{\gamma^d}^d \end{array} \right]~\hbox{Control variables of dynamic buses},\\
v^{nd}\ad =  \left[\begin{array}{c} v_{\gamma^d+1}^{nd}  \\ \vdots \\ v_{\gamma}^{nd} \end{array} \right]~\hbox{Control variables of non-dynamic buses},\\
v \ad =  \left[\begin{array}{c} v^{d} \\  v^{nd} \end{array} \right] ~\hbox{Control variables of all buses},\\
\ell^d\ad =  \left[\begin{array}{c} \ell^d_1 \\ \vdots \\ \ell^d_{\gamma^d} \end{array} \right]~\hbox{Loads of dynamic buses},\\
\ell^{nd}\ad =  \left[\begin{array}{c} \ell^{nd}_{\gamma^d+1} \\ \vdots \\ \ell^{nd}_\gamma \end{array} \right] ~\hbox{Loads of non-dynamic buses},\\
\ell\ad =  \left[\begin{array}{c} \ell^d \\  \ell^{nd} \end{array} \right] ~\hbox{All loads}.
\eea
By (\ref{nd1}),
for non-dynamic buses, we have
\begin{align}
&G^0(z^{nd}, z^d, v^{nd},\ell^{nd})\nonumber\\
=&\left[\begin{array}{c} g_{\gamma^d+1}(z^{nd}_{\gamma^d+1},z_{\gamma^d+1}^-, v_{\gamma^d+1}^{nd},\ell^{nd}_{\gamma^d+1})\\
\vdots\\
g_{\gamma}(z^{nd}_{\gamma},z_{\gamma}^-, v_{\gamma}^{nd}, \ell^{nd}_{\gamma})
\end{array} \right]=0.
\end{align}
For physical power grids, given $z^d$, $v^{nd}$, $\ell^{nd}$, this equation has a unique solution within permitted operating ranges, leading to the symbolic relationship
\beq{nd3}
z^{nd}= H(z^d,v^{nd},\ell^{nd}).
\eeq
Furthermore, by the dynamic systems in (\ref{nd0}),
\beq{d1}
\dot z^d = F^0(z^d, z^{nd},v^d,\ell^d).
\eeq
Substituting (\ref{nd3}) into (\ref{d1}), we obtain
\begin{align}\label{total}
\dot z^d = &F^0(z^d,H(z^d,v^{nd},\ell^{nd}),v^d,\ell^d)\nonumber\\
=&F(z^d,v^{d},v^{nd},\ell^{d},\ell^{nd}).
\end{align}

\textbf{4) Linearization}

In power system control problems, it is common to linearize the nonlinear dynamics (\ref{total})  near nominal operating points \cite{KN,GS}. The linearization process involves the following standard steps. Given the steady-state loads $\overline \ell=[\overline \ell^{d}, \overline \ell^{nd}]^{\top}$ and steady-state input real powers $\overline v=[\overline v^{d}, \overline v^{nd}]^{\top}$, the steady-state
$\overline z^d$ (equilibrium point or the nominal operating condition) is the solution to
$F(\ol z^d,\ol v^d,\ol v^{nd},\overline \ell^{d}, \overline \ell^{nd}) = 0.$

By defining the perturbation variables from their nominal values as
$x=z^{d}-\overline z^{d}, u=v^d-\overline v^d, u^{n}=v^{nd}-\overline v^{nd}, \zeta=\ell^{d}-\overline \ell^{d}, \zeta^n=\ell^{nd}-\overline \ell^{nd}$,
the linearized system is
\beq{total12}
\dot x = A x+B_1 u+ B_2 u^n + D_1 \zeta+ D_2 \zeta^n,
\eeq
where the matrices are  the related Jacobian matrices
\bea
A\ad = \left.{\partial F(z^d,v^{d},v^{nd},\ell^{d},\ell^{nd})\over \partial z^d}\right|_{\tiny{\begin{array}{c} z^d=\ol z^d, v=\ol v, \ell=\ol \ell\end{array}}},\\
B_1\ad = \left.{\partial F(z^d,v^{d},v^{nd},\ell^{d},\ell^{nd})\over \partial v^{d}}\right|_{\tiny{\begin{array}{c} z^d=\ol z^d, v=\ol v, \ell=\ol \ell\end{array}}}, \\
B_2\ad = \left.{\partial F(z^d,v^{d},v^{nd},\ell^{d},\ell^{nd})\over \partial v^{nd}}\right|_{\tiny{\begin{array}{c} z^d=\ol z^d, v=\ol v, \ell=\ol \ell\end{array}}}, \\
D_1\ad = \left.{\partial F(z^d,v^{d},v^{nd},\ell^{d},\ell^{nd})\over \partial \ell^{d}}\right|_{\tiny{\begin{array}{c} z^d=\ol z^d, v=\ol v, \ell=\ol \ell\end{array}}},\\
D_2\ad = \left.{\partial F(z^d,v^{d},v^{nd},\ell^{d},\ell^{nd})\over \partial \ell^{nd}}\right|_{\tiny{\begin{array}{c} z^d=\ol z^d, v=\ol v, \ell=\ol \ell\end{array}}}.
\eea

\begin{rem} The linearization method is commonly used in power system analysis, estimation, and control, in the general framework of small signal analysis.  The main task of linearization involves the computation of Jacobian matrices. This computation can be done either by derivation for small-scale systems or by numerical computation for large-scale systems.

Note that when the operating condition changes, the related Jacobian matrices must be updated due to the change in equilibrium points, leading to a new state space model. Since our linearized model can be easily updated by numerical computations to arrive at new matrices in state space models, our observer designs will naturally adapt to it. As a result,
our algorithms and convergence analysis remain valid in their original form. We emphasized that the issue of stability in adaptation schemes belongs to the adaptive control domain and is out of the scope of this paper, and hence will be left as the subject of future research endeavor.
\end{rem}

\subsection{Illustrative Examples of Model Derivations}

We now use some concrete cases to illustrate the general dynamic modeling approach from the previous subsection. In these examples,  dynamic buses are dispatchable and non-dynamic buses are non-dispatchable, i.e., $v=v^d$, resulting in $B_2=0$.

\textbf{1) Power Flows Between Buses}

\begin{figure}[htb]
	\centerline{\psfig{file=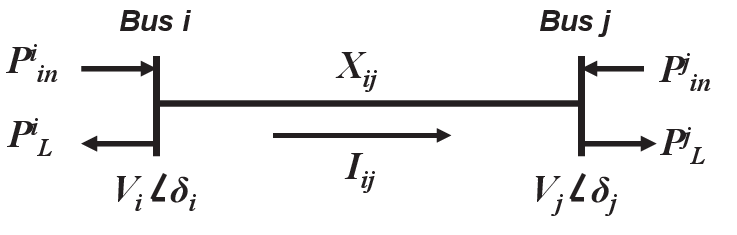,height=1.8cm}}
	\caption{Schematics of a link in microgrids.} \label{connection}
\end{figure}

Real and reactive power flows between buses are governed by the following basic complex power relationship.
Let the voltage of Bus $i$ be denoted by the phasor $\vec V_i=V_i \angle \delta_i$.
For a given link $(i,j)\in \clE$, see Fig. \ref{connection}.

Suppose that the transmission line between Bus $i$ and Bus $j$ has impedance $X_{ij}\angle \theta_{ij}$.   The line current is
 \[
I_{ij} \angle \gamma = {V_i\angle \delta_i - V_j\angle \delta_j \over X_{ij}\angle \theta_{ij}}
 = {V_i\over X_{ij}} \angle (\delta_i -\theta_{ij}) - {V_j\over X_{ij}} \angle (\delta_j-\theta_{ij}).
\]
Denote $\delta_{ij}=\delta_i-\delta_j$.
The complex power flow from Bus $i$ to Bus $j$ at Bus $i$ is
\[S_{ij} = V_i \angle \delta_i \times I_{ij}\angle (-\gamma) =
{V_i^2\over X_{ij}} \angle \theta_{ij} - {V_iV_j\over X_{ij}} \angle (\theta_{ij}+\delta_{ij}),\]
which implies that the transmitted real and reactive powers at Bus $i$ are
\beq{real1} P_{ij} = {V_i^2\over X_{ij}} \cos( \theta_{ij}) - {V_iV_j\over X_{ij}}\cos (\theta_{ij}+\delta_{ij}),
\eeq
\beq{reactive1} Q_{ij} = {V_i^2\over X_{ij}} \sin( \theta_{ij}) - {V_iV_j\over X_{ij}}\sin (\theta_{ij}+\delta_{ij}).
\eeq

\textbf{2) Linearized Dynamic Model of Dynamic Buses}

In frequency regulation, the common dynamic models are swing equations  for synchronous machines near nominal operating conditions or similar types for simplified models for MGs. For Bus $i$, let $\delta_i$ be its electric angle and
$
\omega_i=\dot \delta_i,
$
then the swing equation is
\beq{dy1}
M_i \dot \omega_i +g_i( \omega_i) = P^i_{in}-P^i_L-P^i_{out},
\eeq
where $M_i$ is the equivalent electric-side inertia, $g_i(\cdot)$ represents the nonlinear damping effect.  $g_i(\cdot)$ is continuously differentiable and satisfies $\omega_i g_i(\omega_i)> 0$ for $\omega_i \neq 0$. This implies that $g_i(0)=0$. Since $g_i(0)=0$, its Taylor expansion around $\omega_i=0$ can be written as
$g_i(\omega_i)= b_i \omega_i+\wdt g_i(\omega_i)$,
where the coefficient of the linear term  $b_i > 0$ and $\wdt g_i(\omega_i)$ represents the higher-order terms. Note that $P^i_{in}$ is the equivalent electric-side real power input that is treated as the local control input, and $P^i_L$ is the local real power load on Bus $i$ that is viewed as a disturbance.

\textbf{3) Algebraic Model of Non-Dynamic Buses}

Non-dynamic buses involve only power balance equations. From power flow relationships, $P^i_{out}$ is the total transmitted power from Bus $i$ to its neighboring buses
\begin{align}\label{nondy}
P^i_{out}& = \!\sum_{j\in \clN_i} P_{ij}
\!\nonumber\\
& =\!\sum_{j\in \clN_i} \left[{V_i^2\over X_{ij}} \cos( \theta_{ij}) - {V_iV_j\over X_{ij}}\cos (\theta_{ij}\!+\!\delta_{ij})\right]\nonumber\\
& =\!\sum_{j\in \clN_i} q(\delta_i,\delta_j),
\end{align}
where
$$
q(\delta_i,\delta_j)={V_i^2\over X_{ij}} \cos( \theta_{ij}) - {V_iV_j\over X_{ij}}\cos (\theta_{ij}+\delta_{ij}).
$$

\textbf{4) Integrated Dynamic Models}

All system parameters in (\ref{dy1})  assume constant values under normal operating conditions but may jump to other values under contingency.
Define the local state variable $z^d_i=[\delta_i,\omega_i]^{\top}\in \dbR^2$, the local control $v_i=P^i_{in}$, the local disturbance  $\ell^d_i=P^i_{L}$.  To highlight system interactions, for Bus $i$ we also define the variables of its neighboring buses as $z^-_i$.
  The local dynamics (\ref{dy1}) can be represented by a nonlinear state equation
\beq{dy11}
\dot z^d_i = f_i(z^d_i,z^-_i)+ B^i_1 v_i + D^i_1 \ell_i^{d},
\eeq
where
$$
f_i(z^d_i,z^-_i) =\left[\begin{array}{c} \omega_i \\ - {g_i(\omega_i)\over M_i} -{1\over M_i} \sum_{j\in \clN_i} q(\delta_i,\delta_j) \end{array} \right],
$$
$$
B^i_1= \left[\begin{array}{c} 0 \\ 1/M_i \end{array} \right],~~~D^i_1= \left[\begin{array}{c} 0 \\ -1/M_i \end{array} \right].
$$
Then, denote the local non-dynamic variable as $z_{j}^{nd}=\delta_{j}\in\mathbb{R}$, and
$$
z^d \!=\!\!  \left[\begin{array}{c} z_1^{d} \\ \vdots \\ z_{\gamma^d}^{d} \end{array} \right]\!\!,
z^{nd} \!=\!\!  \left[\begin{array}{c} z_{\gamma^d+1}^{nd} \\ \vdots \\ z_{\gamma}^{nd} \end{array} \right]\!\!,
v \!=\!\! \left[\begin{array}{c} v_1 \\ \vdots \\ v_{\gamma^d} \end{array} \right]\!\!,
$$
$$
\ell^{d} \!=\!\! \left[\begin{array}{c} \ell_1^d \\ \vdots \\ \ell_{\gamma^d}^{d} \end{array} \right]\!\!,
\ell^{nd} \!=\!\!  \left[\begin{array}{c} \ell_{\gamma^d+1}^{nd} \\ \vdots \\ \ell_{\gamma}^{nd} \end{array} \right]\!\!,
F^{0}(z)\!=\!\! \left[\begin{array}{c} f_1(z) \\ \vdots \\ f_{\gamma^d}(z) \end{array} \right]\!\!.
$$

After expressing the variables of non-dynamic buses as functions of the state variables, by (\ref{nd3}) and (\ref{nondy}), we have $z^{nd}= H(z^d,\ell^{nd})$. Combine with (\ref{dy11}), we obtain
\begin{align}\label{dy2}
\dot z^d = & F^{0}(z^{d}, z^{nd})+ B_1 v + D_1 \ell^{d}\nonumber\\
=& F^{0}(z^{d}, H(z^d,\ell^{nd}))+ B_1 v + D_1 \ell^{d}\nonumber\\
= & F(z^{d}, \ell^{nd})+ B_1 v + D_1 \ell^{d},
\end{align}
where $B_1=  \diag[B^i_1], D_1=\diag[D^i_1]$.

\begin{exm}\label{exm0}
Consider a two-bus system shown in Fig. \ref{connection} with $i=1$ and $j=2$. Bus $1$ is a dynamic bus with $v_{1}=P_{in}^{1}, \ell_{1}^{d}=P_{L}^{1}$, and Bus $2$ is a non-dynamic bus with load $\ell_{2}^{nd}=P^2_L$ and no input power $P^2_{in}=0$.
Denote
$$
\begin{aligned}
&\theta=\theta_{12},~~X=X_{12},~~\beta_1= {V_1^2\over X} \cos( \theta),~~\beta_2={V_2^2\over X} \cos( \theta),\\
&\beta=V_1 V_2/X,~~\delta = \delta_1-\delta_2.
\end{aligned}
$$
Define the state variable $z_{1}^d=[\delta_1, \omega_1]^{\top}$, and the non-dynamic variable $z_{2}^{nd}=\delta_{2}$, and assume that
 $g_1(\omega_1)=b_1 \omega_1$, $b_1>0$. Then,
\[
f_1(z_{1}^d, z_{2}^{nd}) = \left[\begin{array}{c} \omega_1 \\ - {b_1 \omega_1 \over M_1} -{1\over M_1}(\beta_1-  \beta\cos (\theta+\delta_1-\delta_2))\end{array} \right].
\]

From the equation on Bus $2$,
$\ell_{2}^{nd}=P^2_L=-\beta_2 + \beta\cos (\theta-\delta)$.
Solving for
$\delta= \theta- \arccos((\beta_2+ P^2_L)/\beta)$
yields that
\bea
\ad f_1^{0}(z_{1}^d, z_{2}^{nd})=f_1(z_1^d, \ell_{2}^{nd}) \\
\ad = \left[\begin{array}{c} \omega_1 \\ - {b_1 \omega_1 \over M_1} -{1\over M_1}\left[\beta_1-  \beta\cos (2\theta-\arccos(\frac{\beta_2+ P^2_L}{\beta}))\right]\end{array} \right].
\eea
The state equation for this grid is
\[ \dot z_{1}^d=f_1(z_1^d, \ell_{2}^{nd})+ \left[\begin{array}{c} 0 \\
1/M_1\end{array} \right] v_1 + \left[\begin{array}{c} 0 \\
-1/M_1\end{array} \right] \ell_1^{d}.
\]
 \end{exm}

\textbf{5) Linearization}

In the state space model (\ref{dy2}),
$\dot z^d=F(z^{d}, \ell^{nd}) + B_1 v +D_1 \ell^d$,
given the steady-state loads $\overline \ell^{d}, \overline \ell^{nd}$, steady-state input real powers  $\overline v$, and the steady-state
$\overline z^d$ (equilibrium point or the nominal operating condition) is the solution to
$$F(\overline z^{d}, \overline \ell^{nd})+B_1 \overline v+ D_1 \overline \ell^d = 0.$$

Define $x=z^d-\overline z^d$, $u=v-\overline v$, $\zeta=\ell^{d}-\overline \ell^{d}$, and $\zeta^{n}=\ell^{nd}-\overline \ell^{nd}$, the linearized system is
\beq{total2}
\dot x = A x+B_1 u+ D_1 \zeta + D_{2}\zeta^{n},
\eeq
where
$$
\begin{aligned}
A=& {\partial F(z^{d}, \ell^{nd})\over \partial z^d}|_{z^d=\overline z^d, \ell^{nd}=\ol \ell^{nd}},\\
D_{2}=& {\partial F(z^{d}, \ell^{nd})\over \partial \ell^{nd}}|_{z^d=\overline z^d, \ell^{nd}=\ol \ell^{nd}}.
\end{aligned}
$$
The following example shows how to linearize the two-bus system of Fig. \ref{connection}.

\begin{exm}\label{exm1}
{\rm Consider the two-bus system shown in Fig. \ref{connection} with $i=1$ and $j=2$, and $\theta_{12}=90^o$. Both buses are generator buses and dynamic.
We have $z_1^d=[\delta_1, \omega_1]^{\top}$,
$z_2^d=[\delta_2, \omega_2]^{\top}$. Suppose that $g_1(\omega_1)=b_1 \omega_1$, $b_1>0$, and
$g_2(\omega_2)=b_2 \omega_2$, $b_2>0$. Denote $\beta=\beta_{12}= V_1 V_2/X_{12}$ and $\delta = \delta_1-\delta_2$. Then,
\bea
f_1(z_1^d,z_2^d)\ad = \left[\begin{array}{c} \omega_1 \\ - {b_1 \omega_1 \over M_1} -{1\over M_1}  \beta\sin (\delta)\end{array} \right],\\
f_2(z_2^d,z_1^d)\ad = \left[\begin{array}{c} \omega_2 \\ - {b_2 \omega_2 \over M_2} -{1\over M_2}  \beta\sin (-\delta)\end{array} \right].
\eea

Given the nominal inputs $\overline v_1=P^1_{in}$, $\overline v_2=P^2_{in}$, and nominal loads $\overline \ell_1^d=P^1_{L}$, $\overline \ell_2^d=P^2_{L}$, we first compute the equilibrium point from
\begin{equation}
\begin{cases}
&\omega_1 =0,\\
 &- {b_1 \omega_1 \over M_1} +{1\over M_1}  (-\beta\sin (\delta) +P^1_{in}-P^1_{L}) =0,\\
 &\omega_2 =0,\\
 &- {b_2 \omega_2 \over M_2} +{1\over M_2}  (\beta\sin (\delta) +P^2_{in}-P^2_{L}) =0.
 \end{cases}
\end{equation}
The solution is $\overline{\omega}_1=0$, $\overline{\omega}_2=0$, and
$\overline \delta = \sin^{-1}\left({P^1_{in}-P^1_{L}\over \beta}\right)$.

The linearized system (\ref{total2}) for this example is
\[ A =\left[\begin{array}{cccc} 0 & 1 & 0 & 0 \\
-{\beta \over M_1}  \cos (\overline \delta) &  - {b_1  \over M_1} & {\beta \over M_1}  \cos (\overline \delta) & 0\\
 0 & 0 & 0 & 1 \\
{\beta \over M_2}  \cos (\overline \delta) & 0 & -{\beta \over M_2}  \cos (\overline \delta) & - {b_2  \over M_2}
 \end{array} \right]\]
 \[ B_1 =\left[\begin{array}{cc}  0 & 0\\
 1/M_1 & 0\\ 0 & 0 \\ 0 & 1/M_2
 \end{array} \right], D_1 =\left[\begin{array}{cc}  0 & 0\\
 -1/M_1 & 0\\ 0 & 0 \\ 0 & -1/M_2
 \end{array} \right].
 \]

 For numerical values,
 assume $M_1=1$, $M_2=1.5$, $b_1=0.2$, $b_2=0.31$, $\beta=200$, $P^1_{in}=100$, $P^2_{in}=50$, $P^1_{L}=70$, $P^2_L=80$.
 Then, the equilibrium point is $\overline\delta = 0.1506$ (rad) and
\[
A =\left[\begin{array}{cccc}
         0   & 1      &   0    &     0\\
 -197.7372  & -0.2 & 197.7372     &    0\\
         0   &      0    &     0  &  1\\
  131.8248    &     0  & -131.8248  & -0.2067
\end{array} \right].
\]
}
 \end{exm}

\section{Contingency and Stochastic Hybrid Systems}\label{Sensors}

Power system contingencies on transmission lines, power generators, control devices, and communication system interruption are random and cause jumps in system structures or parameters in dynamic models. Consequently, they can be modeled as stochastic hybrid systems.
Mathematically, we may list all scenarios of normal operations and contingencies under study as a set $\clS=\{1,\ldots, m\}$ and use a jumping process $\alpha(t) \in \clS$ to represent the occurrence of the corresponding scenario.
In particular, following (\ref{total2}) if we concentrate on linearised systems, they can be generally represented by randomly switched linear systems (RSLSs)
$$
\left\{\begin{array}{rcl}
\dot x &=& A(\alpha(t))x+B_1(\alpha(t)) u+B_2(\alpha(t)) u^n\\
& &+ D_1(\alpha(t)) \zeta+ D_2(\alpha(t)) \zeta^n,\\
y & = & C(\alpha(t)) x.
\end{array}\right.
$$
The dependence of system matrices on contingencies is then represented by their values as functions of $\alpha(t)$.
Since contingencies occur randomly, $\alpha(t)$ is a stochastic process. It should be mentioned that the dimension of the output $y$ may change with $\alpha(t)$.

In general, contingencies can affect all matrices in the system model. However, if we consider only intermittent sensor interruptions, this type of contingency affects only the $C$ matrices. Such contingencies cause jumps in $C$ values and the corresponding RSLSs are simplified to
\beq{sys0}
\left\{\begin{array}{rcl}
\dot x &=& A x+B_1 u+B_2 u^n+ D_1 \zeta+D_2 \zeta^n,\\
y & = & C(\alpha(t)) x,
\end{array}\right.
\eeq
namely, only the output matrix depends on $\alpha(t)$.

Sensor contingencies  cause changes in measured signals and their interruptions.
 The output equation of the system (\ref{sys0}) depends on sensor systems.
This paper considers PMU and frequency data on a limited number of buses. To reduce system and maintenance costs, it is desirable to reduce the number of PMUs while still maintaining reliable state estimation on $x$ so that power system decisions and management can be carried out reliably. The number of PMUs and their locations have an important impact on such reliability analysis.

Sensor systems are reflected in the $C$ matrix in the system model (\ref{sys0}). We list several common sensor configurations. We use \emph{Example \ref{exm1}} for this illustration:
\begin{enumerate}
\item If only $\delta_1$ is measured, then  $y=\delta_1-\overline \delta_1= Cx$, with $C=[1,0,0,0]$.
\item If both  $\delta_1$ (by PMU) and $\omega_1$ (by traditional frequency measurements) are measured, then $C=\left[\begin{array}{cccc} 1 & 0 & 0 &0\\
     0 & 1 & 0 &0\end{array} \right]$.
\item If the PMU measurements on both Bus $1$ and Bus $2$ are used, then  $C=\left[\begin{array}{cccc} 1 & 0 & 0 &0\\
     0 & 0 & 1 &0\end{array} \right]$.
 \end{enumerate}

For state observer  design, the inputs are irrelevant, so we may ignore the inputs to the system and consider a continuous-time RSLS
with output observation noise
\beq{sys}
\left\{
\begin{array}{rcl}\dot x(t) & = & Ax(t),\\
dy(t) & = & C(\alpha(t)) x(t)dt + \sigma(\alpha(t)) d w, \end{array}
\right.
\eeq
where  $x(t)\in \dbR^n$ is the state, $y(t)\in \dbR^r$ is the
output, and $w\in \dbR^r$ is the $r$-dimensional standard real-valued Brownian motion which is independent of $\alpha$ and  $\sigma(i)\sigma(i)^{\top}\geq 0$, $i\in \clS$.
Here, the output equation in (\ref{sys}) is a stochastic differential equation (SDE), that has been ubiquitously used as observation equations in stochastic systems \cite{YinZ10}.
For each given value  $i\in \clS$, the corresponding linear time invariant (LTI) system in (\ref{sys}) with matrices $(C(i), A)$ is called {\it the $i$th subsystem} of the RSLS.
 The following assumption is needed for the analysis.

\begin{asm}\label{asm1} Given a sampling interval $\tau$,
(a) the switching process $\alpha(t)$ can switch only at the instants $k\tau$, $k=0,1,2,\ldots,$ that generates
a random
sequence $\{\alpha_k=\alpha(k\tau)\}$ ({\it the skeleton sequence}). (b) The sequence $\{\alpha_k\}$ is independent and identically distributed  (i.i.d.) with probability ${\rm Pr}\{\alpha_k=i\}=p_i>0$, $i\in \clS$,
and $\sum_{i=1}^m p_i =1$. (c) $\alpha_k$ is independent of $x(0)$ and the Brownian motion $w$. (d) $\alpha(t)$ can be directly measured, but it is not known before its occurrence.
\end{asm}

\begin{rem}
Modeling of random contingencies as switching or jumps in a stochastic hybrid system framework is exemplified
 in various power system contingency scenarios, such as random packet loss in communication channels in  PMUs that use GPS for universal time synchronization and communication systems for data exchange with the system protection center (SPC). Additionally, the framework can  randomly capture sensor failures, circuit breaker trips, line faults, and the activation of surge protection devices.

The time interval $\tau$ depends on hardware/software systems. For example, the PMU  data rate of the Power Xpert Meter is $1024$ samples per cycle. For contingency management,
the IEEE-imposed limit for voltage sag/surge is set at 160 ms, necessitating a data sampling interval significantly shorter than this threshold.
Mathematically, under this assumption, the random switching process can be treated as a discrete-time stochastic sequence, rather than a continuous-time process.
\end{rem}

Under \emph{Assumption \ref{asm1}}, we have $y(t): \dbR^n \to \dbC[0,\tau)$, and the random matrix
sequence
$$
C_k  =C(\alpha_k)=\sum_{i=1}^m C(i) \one_{\{\alpha_k=i\}},
$$
where $\one_G$ is the indicator function of the event $G$: $\one_G =1$ if $G$ is true; and $\one_G =0$, otherwise. Also, $C_k$ is a matrix-valued random variable.
 Let $W$ be the observability matrix of $(C,A)$
\[ W=\left[\begin{array}{c} C\\ C A\\ \vdots \\ C A^{n-1} \end{array} \right].\]

\section{Observer Design for State Estimation under Noisy and Randomly Interrupting Sensor Data}\label{Design}

In this section, we describe observer design procedures. Some of the recently introduced methods from \cite{WY3} are employed.
To be self-contained and to improve
 readability, we include certain derivation details and proofs.

The sampled values of the state are denoted by $x_k=x(k\tau)$ and satisfy
$
x_{k+1}=  e^{A\tau} x_k.
$
The state transition mapping from $x_0$ to $x_k$ is
$x_k=e^{A\tau k} x_0 = H_{k} x_0, k=1,2,\ldots,$
where $H_k=e^{A\tau k}$.

For the $i$th subsystem in $\clS$, the matrices $A$ and $C(i)$  are constant matrices, and their
 observability matrices are
\beq{WWJ} W(i)=\left[\begin{array}{c} C(i)\\ C(i) A\\\vdots \\ C(i) A^{n-1} \end{array} \right],\ i=1,\ldots,m.\eeq
The combined observability matrix for the set $\clS$ is
\beq{Ws} W_\clS= \left[\begin{array}{c} W(1)\\ W(2)\\\vdots \\ W(m) \end{array} \right].\eeq
We note that $W(i)$ and $W_\clS$ are deterministic matrices.

\begin{asm}\label{Ws}
	Assume that (a) subsystems may be unobservable, i.e.,
 $\Rank(W(i))=n_i \leq n$, where $i=1,\ldots, m$.
(b) The combined observability matrix $W_\clS$ is full rank.
\end{asm}

\subsection{Observer Design for Subsystems}
Let $W(i)$ be the observability matrix of the $i$th subsystem defined in (\ref{WWJ}).
If  the $i$th subsystem is unobservable, then
 $\Rank(W(i))=n_i <n$. We construct the base of its kernel as
$M_i=\Base(\ker(W(i))) \in \dbR^{n\times (n-n_i)}$,
 and select any $N_i\in \dbR^{n\times n_i}$ such that
 $T_i=[M_i, N_i]$
 is invertible. The inverse of $T_i$ is decomposed into
 $$T_i^{-1} =\left[\begin{array}{c} G_i\\ F_i \end{array} \right],$$
 where $G_i\in \dbR^{(n-n_i)\times n}$ and $F_i\in \dbR^{n_i\times n}$.

 Define
 $$ \wdt \phi_i =T_i^{-1} x
  =\left[\begin{array}{c} G_i x\\ F_i x \end{array} \right]
  = \left[\begin{array}{c}  \psi^i \\  \phi^i \end{array} \right],$$
 where $  \phi^i \in \dbR^{n_i}$ represents the observable sub-state of the $i$th subsystem.  We focus on constructing a subsystem observer for estimating the sub-state $\phi^i$ when $\alpha_k=i$.

  For the $i$th subsystem, denote
 $   A^i = T_i^{-1} AT_i$, $  C^i = C(i) T_i$, which
 have  the structure
 \[   A^i = \left[\begin{array}{cc}   A^i_{11} &   A^i_{12}\\
 0 &   A^i_{22}\end{array} \right], \quad   C^i= \left[0,   C^i_2\right],\]
 with $  A^i_{22} \in \dbR^{n_i \times n_i}$, $  C^i_{2} \in \dbR^{r \times n_i}$, and  $(C^i_{2},A^i_{22})$ is observable with respect to the sub-state $\phi^i$.

\textbf{1) Dynamics of Errors when $\alpha_k=i$}

 When $\alpha_k=i$,   the open-loop dynamics of $z^i$ is
 \beq{zi}
 \left\{ \begin{array}{rcl} \dot{\phi}^i &=&   A^i_{22}   \phi^i,\\
 d y & =&    C^i_2 \phi^i dt + \sigma_k d w, \end{array}\right.
 \eeq
 and $(C^i_2,   A^i_{22})$ is observable, and $w$ is the $r$-dimensional real-valued standard Brownian motion.
The observer for the $i$th subsystem  in $[k\tau, (k+1)\tau)$  for estimating the sub-state $ z^i$  is
 \beq{hatzi}
  d {\wdh \phi^i} =   A^i_{22} \wdh \phi^i dt+  L_i (dy-  C^i_2 \wdh \phi^i dt),
 \eeq
 where $ L_i \in \dbR^{n_i \times r}$ is the constant observer feedback gain.
 Denote
 $$
 A^i_c=   A^i_{22}-  L_i   C^i_2,~~~e^i=\wdh \phi^i-  \phi^i.
 $$
 Then
 \beq{dei}
 d {e^i}(t)  =  A^i_c e^i(t) dt+  L_i\sigma_k dw(t),
  \eeq
 where $L_i$ is designed such that
 $A^i_c=A^i_{22}-  L_i   C^i_2$ is stable in the continuous-time domain, i.e., all eigenvalues of $A^i_c$ are in the open left half plane.

Consider (\ref{dei}) for $t \in [ k\tau, k\tau+\tau)$ with an initial state $e_k$.
For reliable state estimation, we would like to obtain the distribution of $e^i_k=e^i(k\tau)$. Since $e^i_k$ is Gaussian distributed, its expectation $m^i_k=\mathbb{E}(e_k)$ and variance $V_k=\mathbb{E}((e_k-m_k)(e_k-m_k)^{\top})$  completely characterize the distribution of $e_k$.

\begin{lem}\label{var}\cite{WY3}
Suppose that $e^i(t)$ is a solution of the stochastic differential equation given by \eqref{dei} together with initial data $e^i_k$ and $A^i_c$ is Hurwitz, i.e.,  all of its eigenvalues are in the open left half plane of the complex plane. Then we have
\begin{equation}\label{mk}
m_{k+1}^{i}=e^{A^i_c \tau}m_k^{i},
\end{equation}
and
\begin{equation}\label{Vk}
V_{k+1}^{i}= \int^\tau_0 e^{A^i_c \mu} L_i\sigma_k (L_i\sigma_k )^{\top} e^{(A^i_c)^{\top}\mu} d\mu.
\end{equation}
\end{lem}

Based on \emph{Lemma \ref{var}}, denote $Q^i_{k}= (V^i_{k+1})^{1/2}$. Since $e^i_{k+1}$ is Gaussian, which is completely determined by $m_{k+1}^{i}$ and $Q^i_k$, by \emph{Lemma \ref{var}},  $e^i_{k+1}$ can be expressed as
\beq{eki} e^i_{k+1} = e^{A^i_c\tau}e_k + Q^i_k d_k,\eeq
where $d_k$ is an i.i.d. sequence and $d_k\sim N(0,I_{n_i})$.

\textbf{2) Dynamics of Errors when $\alpha_k=j\neq i$}

 Define $n_s= \sum_{i=1}^m n_i$ and
 $$
 F=\left[\begin{array}{c} F_1\\ \vdots \\ F_m \end{array} \right]\in \dbR^{n_s\times n},
 $$
 which is full rank under \emph{Assumption \ref{Ws}}.
From subsystem observers,
  define
  $$ \phi=\left[\begin{array}{c}   \phi^1\\ \vdots\\   \phi^m \end{array} \right]\in \dbR^{n_s},~~~~
  \wdh \phi=\left[\begin{array}{c} \wdh \phi^1\\ \vdots\\ \wdh \phi^m \end{array} \right]\in \dbR^{n_s}.$$
  Then,
  $\phi=Fx$ and $\wdh \phi=F\wdh x.
  $
   Define $\Phi= (F^{\top}F)^{-1} F^{\top}\in \dbR^{n\times n_s}$ is of (row) rank $n$ , and
  $x=\Phi \phi$. Consequently,
  their sampled values are
  $ x_k= \Phi   \phi_k$, $\wdh x_k= \Phi \wdh \phi_k.$

  Since the true system is $x_{k+1}=e^{A\tau} x_k$, the true sampled value of the subsystem state is
  $$
  \phi^i_{k+1} = F_i x_{k+1} = F_i e^{A\tau} x_k.
  $$
  Consequently, when $\alpha_k\neq i$, the subsystem observer runs open-loop with
 $$
 \wdh \phi^i_{k+1} = F_i \wdh x_{k+1} = F_i e^{A\tau} \wdh x_k,  \ \text{ if } \ \alpha_k=j\neq i.
 $$
  The observer for estimating $x(t)$ is $\wdh x_k=\Phi \wdh \phi_k$. Denote the estimation errors as $\e_k= \wdh x_k - x_k$ and $e_k= \wdh \phi_k- \phi_k$. We have $\e_k= \Phi e_k$ and
  \begin{equation}\label{ekj}
  e^i_{k+1}=F_ie^{A \tau} \e_k = F_ie^{A \tau} \Phi e_k, \ \text{ if } \ \alpha_k=j\neq i.
  \end{equation}

  \textbf{3) Dynamics of Total Error}

    For the $i$th subsystem, by (\ref{eki}) and (\ref{ekj}), we know that
  $$
  e^i_{k+1} = \left\{\begin{array}{cl}
  e^{A^i_c \tau} e^i_k + Q_k^{i} d_k, & \hbox{if~~ $\alpha_k=i$}\\
  F_ie^{A \tau} \Phi e_k,& \hbox{if~~ $\alpha_k\neq i$}
  \end{array}
  \right.
  $$ and
  \bea
  \ad e_{k+1}  =\left[\begin{array}{c} e^1_{k+1}\\ \vdots\\ e^m_{k+1} \end{array}\right]\\ \\
 =\ad \left[\begin{array}{c} \one_{\{\alpha_k=1\}} (e^{A^1_c \tau} e^1_k\! +\!Q^1_k d_k)\! +\! \one_{\{\alpha_k\neq 1\}}F_1e^{A \tau} \Phi e_k\\ \vdots\\
  \!\!\one_{\{\alpha_k=m\}}(e^{A^m_c \tau} e^m_k\!+\!Q^m_k d_k) \!+\! \one_{\{\alpha_k\neq m\}}F_me^{A \tau} \Phi e_k \!\! \end{array} \right]\\
  \\
  =\ad \Lambda^1 e_k+\Lambda^2 F e^{A\tau}\Phi e_k+\Gamma_k d_k,
    \eea
    where
  $
  \Lambda^1=\diag[\one_{\{\alpha_k=1\}} e^{A^1_c \tau},\ldots,\one_{\{\alpha_k=m\}} e^{A^m_c \tau}],
  $
  $
  \Lambda^2=\diag[\one_{\{\alpha_k\neq 1\}},\ldots,\one_{\{\alpha_k\neq m\}}],
  $
  and
  $$
  \Gamma_k=\left[\begin{array}{c}\one_{\{\alpha_k=1\}} Q^1_k\\ \vdots\\ \one_{\{\alpha_k=m\}} Q^m_k\end{array}\right].
  $$
  By defining $\Lambda_k=\Lambda^1 +\Lambda^2 F e^{A\tau} \Phi$, we obtain the total error dynamics
  \beq{LAM} e_{k+1}=\Lambda_k e_k + \Gamma_k d_k.\eeq

The observer structure for subsystems, the combined observer for $x$, and estimation error dynamics are established here.
To achieve convergence, the error dynamics must be stable. This stability condition depends on the design of $L_i$.

\begin{rem}  We should emphasize that
during implementation,  the observers for all subsystems are not  executed concurrently.
This paper focuses on the continuous-state estimation of $x$. Only the estimated $\hat x(t)$ is kept at the end of each time interval, say $k\tau_-$. At $k\tau_+$, $\alpha_k$ occurs, say $\alpha_k=i$, which is assumed known in this paper (detection of $\alpha_k$ is treated  in a separate paper of ours).
As a result, the $i$th observer is implemented. The estimate $\hat x(k\tau)$ is mapped to the $i$th subsystem as the initial state, and the observable sub-state $z^i$ is estimated and the unobservable substate $v^i$ is running open-loop using the system model. At $(k+1)\tau_-$, both $\hat z^i((k+1)\tau_-)$ and $\hat v^((k+1)\tau_-)$ are re-combined and mapped back to be $\hat x ((k+1)\tau_-)$. The process re-starts for the next interval.

Furthermore, for different contingencies, the observable and unobservable sub-states are very different. As a result, the observable sub-state $z^i$ is only active when $\alpha_k=i$.
One of the main challenges of this paper is to devise an algorithm that can combine subsystem processes into an estimation process for $x$ and also suitable design methods for subsystem observers so that the entire state estimation can be convergent.
\end{rem}

\subsection{Observer Pole Placement Design}

From the error dynamics (\ref{LAM}), the solution is
$$
e_k= \left(\prod_{j=0}^{k-1} \Lambda_{j} \right) e_0 + \sum_{j=0}^{k-1} \Lambda_{k-j} \Gamma_j d_j.
$$
Stability of the error dynamics and convergence of the error variances to their steady-sate values rely on the condition $\gamma = \mathbb{E}(\|\Lambda_k\|) < 1$. The goal of design is to ensure that $\gamma <1$.

\begin{thm}\label{thm1}
	Under \emph{Assumptions \ref{asm1}} and \emph{\ref{Ws}}, there exists $\tau_{\max} > 0$ such that for any $\tau< \tau_{\max}$, $L_i$, $i=1,\ldots,m$,  can be designed such that $\gamma <1$.
\end{thm}

\begin{IEEEproof}
	Since the poles of $e^{A^i_c \tau}$ can be arbitrarily placed, $\gamma^i=\|e^{A^i_c\tau}\|$ can be made arbitrarily small.
	As a result,
	$\gamma_{\max} =\max_{i=1,\ldots,m} \gamma^i$ can be made arbitrarily small.
	Denote $p_{\max}= \max_{i=1,\ldots,m} \{p_i\}$, $q_{\max}= \max_{i=1,\ldots,m} \{1-p_i\}$.
	By \emph{Assumption \ref{asm1}}, $0 < p_{\max} < 1$ and $0 < q_{\max} < 1$.
	Consequently, subsystem observers can be designed to satisfy
	\beq{gamma1} \|\diag[p_1 e^{A^1_c \tau},\ldots,p_m e^{A^m_c \tau}]\|\leq p_{\max} \gamma_{\max} = \gamma_1 < \varepsilon, \eeq
	for any $\varepsilon >0$ .
	
	On the other hand, the operator norm (the largest singular value)
	\[h(\tau) = \|F e^{A\tau} \Phi\|=\|F e^{A\tau} (F^{\top}F)^{-1} F^{\top}\|\]
	is a continuous function of $\tau$ and for $\tau=0$
	\bea h(0)\ad =\|F (F^{\top}F)^{-1} F^{\top}\|\\
	\ad = \lambda_{\max} (F(F^{\top}F)^{-1} F^{\top})\\
	\ad = \lambda_{\max} ((F^{\top}F)^{-1} F^{\top}F) \hbox{  by Jacobson's Lemma}\\
	\ad = \lambda_{\max} (I_s)\\
	\ad= 1,
	\eea
	where $\lambda_{\max}$ is the largest eigenvalue. Since $q_{\max} < 1$, by continuity of $h(\tau)$, there exists $\tau_{\max}>0$ such that
	\begin{align}\label{Tdesign}
	\gamma_2=&\max_{\tau< \tau_{\max}}\|\diag[q_1,\ldots,q_m] F e^{A\tau} \Phi\|\nonumber\\
	\leq &q_{\max} h(\tau_{\max}) < 1-\varepsilon,
	\end{align}
	for some $\varepsilon >0$.
	
	Together, by selecting  $\tau$ first to satisfy $\tau<\tau_{\max}$ in  (\ref{Tdesign}), and then designing subsystem observers to satisfy (\ref{gamma1}),
	we obtain
	$\gamma\leq \gamma_1 + \gamma_2 < \varepsilon + 1-  \varepsilon =1.$
\end{IEEEproof}

\emph{Theorem \ref{thm1}} ensures  that as long as $\tau < \tau_{\max}$, we can always design $L_i$ by the pole placement method so that $\gamma < 1$. This is achieved by first choosing pole positions for the subsystem observer's error dynamics suitably and then designing $L_i$ to achieve the selected pole positions. We will use the case study in the next section to illustrate the design steps and discuss tradeoffs.

\subsection{Steady-State Estimation Error Variances}

We now study the steady-state estimation error from the error dynamics (\ref{LAM}), and discuss some design tradeoffs that are inherent between stability and steady-state error variances. Denote
$
W_k = \mathbb{E}_{\alpha_k} (e_ke_k^{\top}), \mu_k =  \Tr(W_k).
$
 Select $\tau< \tau_{\max}$ and design $L_i$ so that the matrix
 $
 M=\mathbb{E}(\Lambda^{\top}_k\Lambda_k)>0,
 $
 is stable in the discrete-time domain, namely all of its eigenvalues are inside the open unit circle. Let $S=M^{1/2}>0$ be the matrix square root of $M$. Then $M=S^{2}$. Denote $\Psi=\mathbb{E}(\Gamma^{\top}_k \Gamma_k)$.

\begin{thm}
Under \emph{Assumptions \ref{asm1}} and \emph{\ref{Ws}}, the steady-state variance $V_\infty$  satisfies
\begin{equation}\label{mui}
\mu_\infty=\Tr (W_\infty),
\end{equation}
where $W_\infty$ satisfies the Lyapunov equation
$$SW_\infty S - W_\infty + \Psi =0.$$

\end{thm}

 \begin{IEEEproof}
 Since $\Lambda_k$ and $\Gamma_k$ are induced by $\alpha_k$, they are i.i.d. and $M$ and $\Psi$ are constant matrices.
 By \emph{Assumption \ref{asm1}}, $\Lambda_k$, $e_k$, and $d_k$ are mutually independent. Consequently, noting that $\Tr$ and $\mathbb{E}$ commute,
 \bea
 \mu_{k+1}\ad = \Tr(W_{k+1})\\
 \ad = \Tr (\mathbb{E}(\Lambda_k e_k e^{\top}_k \Lambda^{\top}_k)) + \Tr (\mathbb{E}(\Gamma_k d_k d^{\top}_k\Gamma^{\top}_k))\\
 \ad = \mathbb{E}(\Tr(\Lambda_k e_k e^{\top}_k \Lambda^{\top}_k)) + \mathbb{E}(\Tr(\Gamma_k d_k d^{\top}_k\Gamma^{\top}_k))\\
 \ad = \Tr(\mathbb{E}(\Lambda^{\top}_k\Lambda_k e_k e^{\top}_k)) + \Tr(\mathbb{E}(\Gamma^{\top}_k\Gamma_k d_k d^{\top}_k))\\
 \ad = \Tr (M W_k  + \Psi)\\
 \ad = \Tr (SW_k S+ \Psi).
 \eea
 Thus, if the steady-state variance $W_\infty$ solves
    the Lyapunov equation
\beq{sif} SW_\infty S - W_\infty + \Psi =0,\eeq
then $\mu_\infty=\Tr (W_\infty)$.
\end{IEEEproof}

\smallskip

There is an inherent tradeoff in designing $L_i$. By the expression of $\Gamma_k$, it is clear that the larger the feedback gains, the larger the steady-state error $\mu_\infty$.  On the other hand, to increase convergence speed, it is desirable to place poles with larger absolute values of their negative real parts. In other words,
fast convergence and small steady-state errors are in conflict. We will use the case study to explain the tradeoff and discuss some potential approaches for choosing suitable design parameters.

\section{Evaluation Case Studies}\label{case}

\subsection{IEEE 5-Bus System}

In this subsection, we use the IEEE 5-bus system to demonstrate our design procedures and their evaluations.

\textbf{1) Modeling and Linearization of the IEEE 5-Bus System}

In dynamic modeling, if we concentrate on real power management for frequency regulation and assume that the real and reactive power decoupling, then only real powers from generators and loads are involved. Voltage regulations and reactive power control are viewed as external to the problem.
Although voltage magnitudes are assumed to be constants under normal operating conditions, under contingency scenarios, such as faults in voltage management, excitation systems, or transmission lines, there can be jumps in values of bus voltages and other parameters.

For the modeling of the IEEE 5-bus system, we assume that	(a) The  dynamic buses contain suitable voltage control systems so that bus voltage magnitudes are treated as constants under normal operating conditions;
(b) the dynamic buses contain both controllable real power input $P$ and uncontrolled real power load $P_L$;  (c) the non-dynamic buses have uncontrolled real power loads $P_L$ but no controllable real powers.

 It is noted that our state model framework on hybrid systems is general and scalable, and can cover different scenarios of power systems and all bus types.
The case study of the IEEE 33-bus system in the next subsection will use common assumptions on buses, such as PQ buses for loads in which both voltage magnitudes and angles are variables.

The IEEE 5-bus system, shown in Fig. \ref{5bus}, has been widely used in performing power system analysis, performance evaluation, and contingency detection.
In consideration of renewal generation situations, we designate Bus 1 and Bus 2 as  dynamic buses with generators. Their bus  voltage magnitudes are controlled to their rated values by Var condensers or compensators.  Similarly, in light of rapid advancement in Var compensation technology such as flexible AC transmission systems (FACTS), we assume that other load buses have their voltage magnitudes maintained near the rated values. This assumption will be removed in the IEEE 33Bus case studies.

\begin{exm}\label{5b}
{\rm Consider the IEEE 5-bus system shown in Fig. \ref{5bus}. The bus structure and data are from the open-source information in \cite{Tan}. Bus 1 and Bus 2 are dynamic buses and Buses 3-5 are non-dynamic buses.
\begin{figure}[htb]
\centerline{\psfig{file=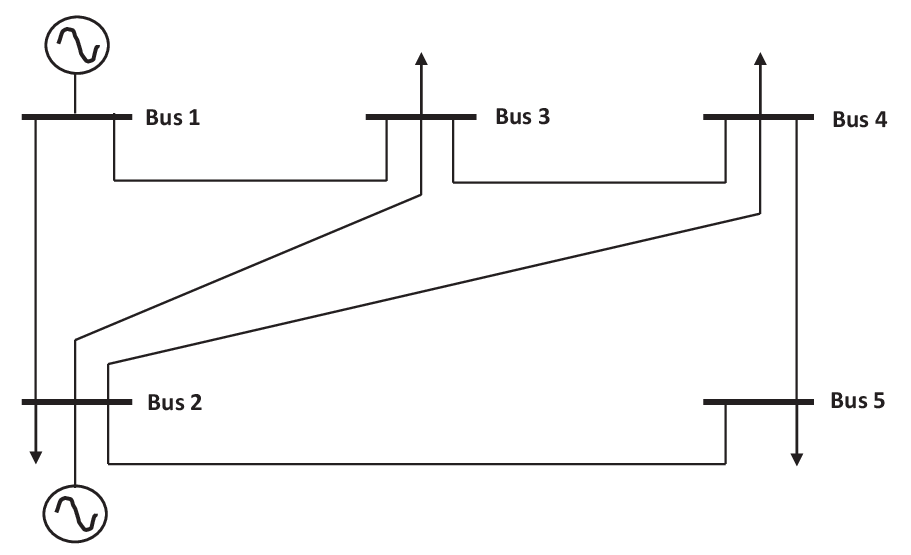,height=4cm}}
\caption{IEEE 5-bus system.} \label{5bus}
\end{figure}

The most common dynamic buses are synchronous generators \cite{K}. Denote
$$
\omega_1=\dot \delta_1, ~~z_1^d=[\delta_1, \omega_1]^{\top}, ~~\omega_2=\dot \delta_2,~~ z_2^d=[\delta_2, \omega_2]^{\top}.
$$
The dynamic systems are
\bea M_1 \dot \omega_1 + g_1(\omega_1)\ad = P^1_{in} - P^1_L+ P^1_{21}+P^1_{31},\\
M_2 \dot \omega_2 + g_2(\omega_2)\ad = P^2_{in} - P^2_L+ P^2_{12}+P^2_{32}+ P^2_{42}+P^2_{52},
\eea
where  the real power flow from Bus $i$ into Bus $j$ on Bus $j$ is
\beq{Pij}P^j_{ij}= {V_j^2\over |Z|_{ij}} \cos( \theta_{ij}) - {V_iV_j\over X_{ij}}\cos (\theta_{ij}+\delta_{ij}),\eeq
and $\delta_{ij} =\delta_i-\delta_j$. The damping term $g_i(w_i)$ has the linear part $b_i \omega_i$ with $b_i>0$, $i=1,2$. The three load buses have real-power equations
$$
\begin{cases}
&P^3_L  = P^3_{13}+P^3_{23}+ P^3_{43},\\
&P^4_L  = P^4_{24}+P^4_{34}+ P^4_{54},\\
&P^5_L  = P^5_{25}+P^5_{45},
\end{cases}
$$
where $P^j_{ij}$ are given in (\ref{Pij}).
Denote
$
z^{d}=[z_1^d, z_2^d]^{\top}, z^{nd}=[\delta_3, \delta_4, \delta_5]^{\top}$. The dynamic systems  can be expressed as a nonlinear state equation
\begin{align}
\dot z^d = & F^{0}(z^d, z^{nd}) + B_1 v+D_1 \ell^{d}\nonumber\\
= & F(z^d, \ell^{nd}) + B_1 v+D_1 \ell^{d},
\end{align}
where
$
v=\left[\begin{array}{c} P^1_{in} \\
P^2_{in} \end{array} \right], ~~\ell^{d}=\left[\begin{array}{c} P^1_L \\
P^2_L\end{array} \right],~~\ell^{nd}=\left[\begin{array}{c} P^3_L \\
P^4_L \\ P^5_L\end{array} \right],$
$$
B_1=\left[\begin{array}{cc} 0&0  \\
1/M_1 &0 \\
0&0  \\
0&1/M_2
\end{array} \right], ~~D_1=\left[\begin{array}{cc} 0&0  \\
-1/M_1 &0 \\
0&0  \\
0&-1/M_2
\end{array} \right].
$$
Denote the perturbations from the nominal values as
$x=z^{d}-\overline z^{d}, u=v-\overline v, \zeta=\ell^{d}-\overline \ell^{d}, \zeta^n=\ell^{nd}-\overline \ell^{nd}$. By (\ref{total2}), the dynamic systems can be linearized near the nominal operating points as
\beq{IEEEfive}
\dot x = A x+B_1 u+ D_1 \zeta + D_2 \zeta^{n},
\eeq
where the matrices are  the related Jacobian matrices
\bea
A\ad = \left.{\partial F(z^d, \ell^{nd})\over \partial z^d}\right|_{\tiny{\begin{array}{c} z^d=\ol z^d, \ell^{nd}=\ol \ell^{nd}\end{array}}},\\
D_2\ad = \left.{\partial F(z^d, \ell^{nd})\over \partial \ell^{nd}}\right|_{\tiny{\begin{array}{c} z^d=\ol z^d, \ell^{nd}=\ol \ell^{nd}\end{array}}}.
\eea
The bus line parameters are extracted from \cite{Tan} and listed in Table \ref{lines}, with $R$ = Resistance, $X$ = Reactance, and $Z=|Z|\angle \theta $ = Impedance.
\begin{table}[h!]
\centering
\caption{IEEE 5-Bus System Line Parameters}
\label{lines}
\begin{tabular}{|c|c|c|c|} \hline
 Line &  $R$ (pu)   &  $X$ (pu) & Z (p.u $|Z|\angle \theta $ rad)   \\ \hline
1-2 & 0.02  & 0.06 & $0.06   \angle 1.25$ \\
1-3 & 0.08  & 0.24 & $0.25  \angle 1.25$ \\
2-3  & 0.06  &  0.25 & $0.26 \angle  1.33$ \\
2-4  & 0.06  & 0.18 & $0.19  \angle 1.25$ \\
2-5 & 0.04  & 0.12 & $0.13 \angle  1.25$ \\
3-4 & 0.01  & 0.03 & $0.03 \angle  1.25$ \\
4-5 &  0.08  &  0.24 & $0.25\angle  1.25$ \\ \hline
\end{tabular}
\end{table}

The nominal operating condition defined in \cite{Tan,Simudata}  is used here with the nominal bus voltages, generation powers and load powers listed in Table \ref{buses} with real power  $P$ (MW) and reactive power $Q$ (MVar). The base MVA is $S_B= 100$ MVA  and the base voltage is $V_B= 230$ kV.

\begin{table}[h!]
\centering
\caption{IEEE 5-Bus System Bus Data}
\label{buses}
{\small
\begin{tabular}{|c|c|c|c|c|c|} \hline
Bus  & $V$ (pu $\angle$ rad)  &  $P$  & $Q$  &  $P_L$  & $Q_L$   \\ \hline
1 & $1.06 \angle 0$ & $129$ & $-7.42$ &  $0$ & $0$ \\
2 & $1.0474 \angle -2.8063$  & $40$ & $30$ & $20$ & $10$ \\
3  & $1.0242 \angle -4.997$   & $0$ & $0$ & $45$ & $15$  \\
4  & $1.0236 \angle -5.3291$   & $0$ & $0$ & $40$ & $5$ \\
5 & $1.0179\angle -6.1503$   & $0$ & $0$ & $60$ & $10$  \\ \hline
\end{tabular}
}
\end{table}

Under the per unit system, the normalized generator parameters are $M_1=1.9$ and  $b_1=0.2$ with equivalent time constant $T_1=M_1/b_1=9.5$ second for Generator 1, and  $M_2=0.9$, $b_1=0.16$ with equivalent time constant $T_2=M_2/b_2=5.625$ second for Generator 2, see \cite{Simudata} for the computational results. Under the aforementioned operating conditions, we obtain
\[A=\left[\begin{array}{cccc}      0  &   1 &         0   &      0\\
    7.7926 &  -0.1053 &  -7.7926   &      0\\
         0     &    0     &    0   & 1\\
  -20.3866     &    0  & 20.3866  & -0.1778
\end{array}\right].
\]
Since eigenvalues of $A$ are $\{-5.3880, 5.2302, 0, -0.1253\}$ which contain a positive value, it is an unstable system.
}
 \end{exm}

\textbf{2) Derivation of the  RSLS for the IEEE 5-Bus System}

We now use the IEEE 5-bus system in \emph{Example \ref{5b}} to illustrate how to derive the RSLS under selected sensor system configurations.

\begin{exm}\label{5bussensor}
{\rm Consider the IEEE 5-bus system in \emph{Example \ref{5b}}. In energy management systems  of a power grid, the voltage angles on some selected buses can be measured by PMUs. Similarly, frequencies at some selected buses are typically measured for frequency regulations.

(i) Sensor systems and observability:

Each sensor represents an output equation. For cost reduction and maintenance simplification, it is highly desirable to reduce sensor complexity. Since different sensor systems have their own inherent observability properties, one may select sensor systems with better observability in our case.

\begin{table}[h!]
\centering
\caption{Sensor Systems and Observability}
\label{sen}
\begin{tabular}{|c|c|c|c|c|c|} \hline
$S_1$  & $S_2$ &  Normal  &  $S_1$ Fails  &  $S_2$ Fails & Both Fail  \\ \hline
$\delta_1$ & $\delta_2$ & Yes & Yes &  Yes & No \\ \hline
$\delta_1$ & $\omega_1$ & Yes & No &  Yes & No \\ \hline
$\omega_1$ & $\omega_2$ & No & No &  No & No \\ \hline
$\delta_1$ & $\omega_2$ & Yes & No &  Yes & No \\ \hline
\end{tabular}
\end{table}

Table \ref{sen} lists some common sensor systems and their observability under normal and contingency conditions. Let $S_i=$ the $i$th sensor, $i=1,2$; Yes $=$ Observable; No $=$ Unobservable. Since the sensor system with $(\delta_1,\delta_2)$ has better observability property, we will use this as an example here.

(ii) Contingency and stochastic switching:

In modern microgrids, the measured signals are communicated to the supervisory control and data acquisition (SCADA) and EMS via wireless communication networks.
Due to communication packet loss, the packet delivery ratio for each channel is given by $\rho_j>0$ for successful data transfer on channel $j$.

Suppose that  $\delta_1$ and $\delta_2$ are independently measured. By adding measurement and communication noises in PMU measurement errors \cite{PMU1,PMU2}, we have $C(1)= \left[\begin{array}{cccc} 1 & 0 & 0 & 0\\  0 & 0 & 1 & 0\end{array}\right]$ for normal operation, $C(2)=[1,0,0,0]$ for failure of Sensor 2 ($\delta_1$ measurement only), $C(3)=[0,0,1,0]$ for failure of Sensor 1 ($\delta_2$ measurement only), and $C(4)=[0,0,0,0]$ for failure on both sensors. Suppose that the packet delivery ratio for Sensor 1 is $\rho_1=0.99$ and for Sensor 2 is $\rho_2=0.995$. This data acquisition scheme can be modeled by an i.i.d. stochastic process $\alpha_k \in \clS=\{1,\ldots,4\}$ with $p_1=\rho_1 \rho_2=0.98505$, $p_2=\rho_1 (1-\rho_2)=0.00495$, $p_3=(1-\rho_1) \rho_2=0.00985$, $p_4=(1-\rho_1) (1-\rho_2)=0.00005$.
}
\end{exm}

\textbf{3) Performance Evaluations and Other Related Issues}

(i) Observer design and convergence:

The pole placement design is used for designing observer feedback gains for $\alpha_k=1$, $\alpha_k=2$ and $\alpha_k=3$. Since $C(4)=0$, the observer can only run open-loop.
For example, if we choose the desired closed-loop poles as $\lambda=[-4.8,-3.6,-4,-4.4]$, then the Matlab function $L_i= place(A',C'(i), \lambda)$, $i=1,2,3 ,4$, yields the suitable feedback gains and the closed-loop error dynamics with
$A_c^i=A-L_i C(i)$, $i=1,2,3 ,4$.

For this example, the upper bound on $\tau$ can be computed as $\tau_{\max}= 0.7365$ (second), which is a sufficient condition for a feasible observer design. Select $\tau = 0.6261$ (second). Denote $M(i)= e^{A_c^i \tau}$, $i=1,2,3,4$.
The closed-loop error dynamics of $e_k$  is a stochastic system with noise-free dynamics $e_{k+1}=M_k e_k$, where
$M_k = \sum_{j=1}^4 \one_{\{\alpha_k=j\}} M(i)$.

\begin{figure}[htb]
	\setlength{\unitlength}{0.1 in}
	\begin{center}
		\includegraphics[height=5.5cm]{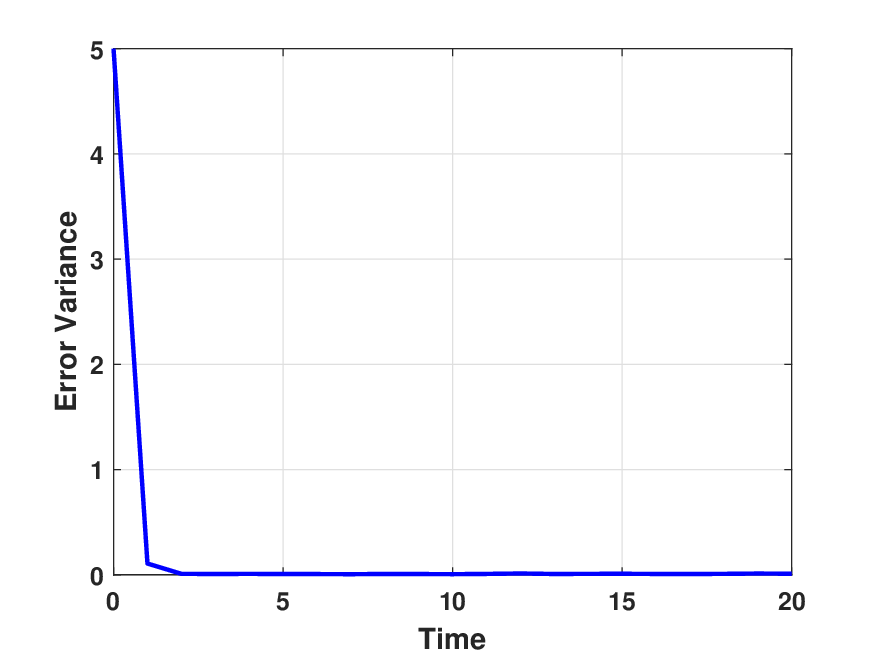}
		\caption{State estimation error variance trajectories.}
		\label{error2}
	\end{center}
\end{figure}

The initial estimation error is selected to be $e(0)=[2,0,1,0]^{\top}$ with the square of error norm $5$.
Suppose that the standard deviations are $\sigma(1)=0.01 I_2$;
$\sigma(2)=0.0015$; $\sigma(3)=0.002$.
Fig. \ref{error2} is the estimation error variance trajectory that  shows that the estimation error is convergent.

\begin{figure}[htb]
	\setlength{\unitlength}{0.1 in}
	\begin{center}
		\includegraphics[height=5.5cm]{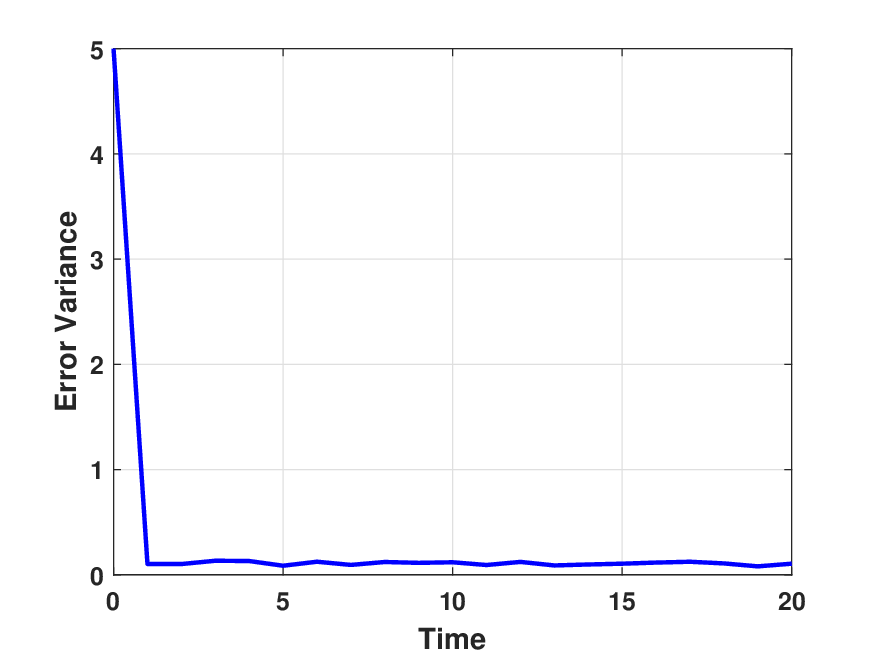}
		\caption{State estimation error variance trajectories under a more aggressive observer design.}
		\label{error3}
	\end{center}
\end{figure}

(ii) Design tradeoff:

We now illustrate the design tradeoff discussed in Section \ref{Design}. Suppose that we now select the designed pole positions to be more negative at $[-9.6,-7.2,-8,-8.8]$, which represents a more aggressive observer design,
and the estimation error converges faster. However, the steady-state errors become bigger due to much larger feedback gains.
Fig. \ref{error3} shows estimation error variance trajectories, showing larger persistent errors.

\begin{figure}[htb]
\setlength{\unitlength}{0.1 in}
\begin{center}
\includegraphics[height=5.5cm]{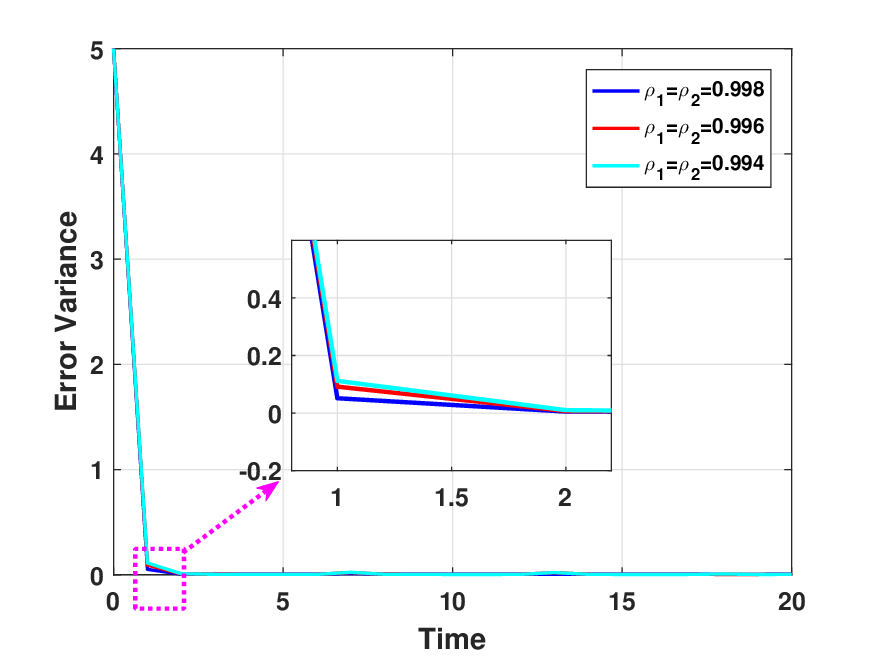}
\caption{State estimation error variance trajectories under different packet delivery ratios.}
\label{errorRe1}
\end{center}
\end{figure}

(iii) Impact of sensor system reliability:

The reliability of the sensor affects state estimation errors.
We choose the desired closed-loop poles to be $\lambda=[-3.6,-2.7,-3,-3.3]$. Consider the packet delivery ratio for Sensor 1 and Sensor 2 in the following four cases:
\begin{table}[h!]
\centering
\caption{The packet delivery ratio}
\label{packet}
\begin{tabular}{|c|c|c|c|c|c|} \hline
 & Case 1 &  Case 2   & Case 3  &  Case 4  \\ \hline
$\rho_1$ & 0.998 & 0.996 & 0.994 &  0.8 \\ \hline
$\rho_2$  & 0.998 & 0.996 & 0.994 &  0.8\\ \hline
\end{tabular}
\end{table}
Fig. \ref{errorRe1} shows estimation error variance trajectories in Case 1-3.
It can be seen that a larger packet delivery ratio leads to 
a faster convergence.

\begin{figure}[htb]
\setlength{\unitlength}{0.1 in}
\begin{center}
\includegraphics[height=5.5cm]{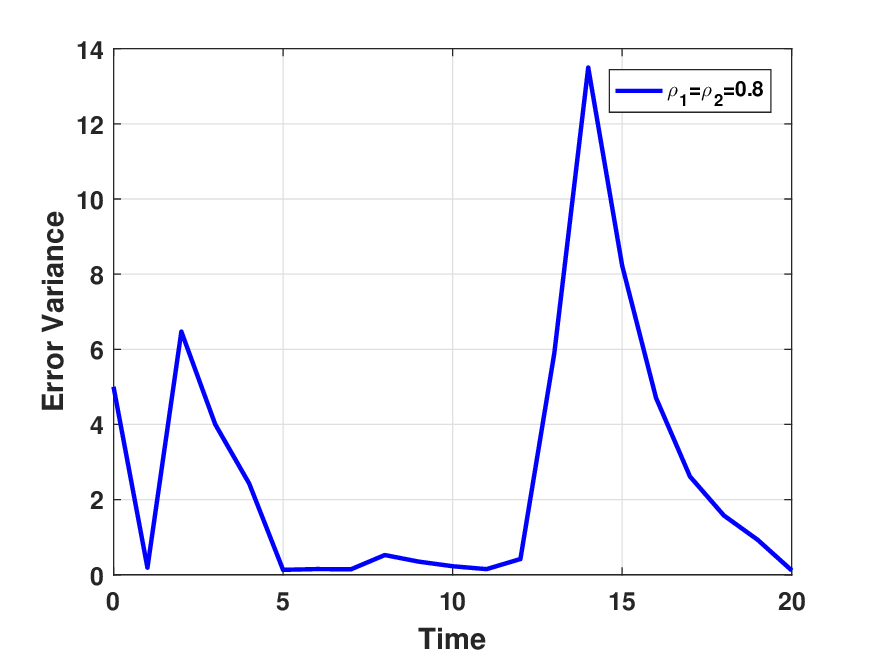}
\caption{State estimation error variance trajectories under Case 4.}
\label{errorRe2}
\end{center}
\end{figure}

Fig. \ref{errorRe2} shows that a small packet delivery ratio may cause the estimator to lose effectiveness. 

(iv) Impact of sensor selections:

Sensor selection  is a very important step of our framework and methods. 
We now use an example to illustrate the impact of sensor selection. 
It is noted that measuring $\delta_1$ and $\delta_2$ means that two PMUs are placed on two different buses. We now consider a less expensive sensor system that measures $\delta_1$ and $\omega_1$ on Bus 1 only. This system uses only one PMU and the frequency measurement is typically in the system already for frequency regulations. In this sensor configuration, we have
$C(1)= \left[\begin{array}{cccc} 1 & 0 & 0 & 0\\  0 & 1 & 0 & 0\end{array}\right]$ for normal operation, $C(2)=[1,0,0,0]$ for failure of Sensor 2 ($\delta_1$ measurement only), $C(3)=[0,1,0,0]$ for failure of Sensor 1 ($\omega_1$ measurement only), and $C(4)=[0,0,0,0]$ for failure on both sensors.

\begin{figure}[htb]
\setlength{\unitlength}{0.1 in}
\begin{center}
\includegraphics[height=5.5cm]{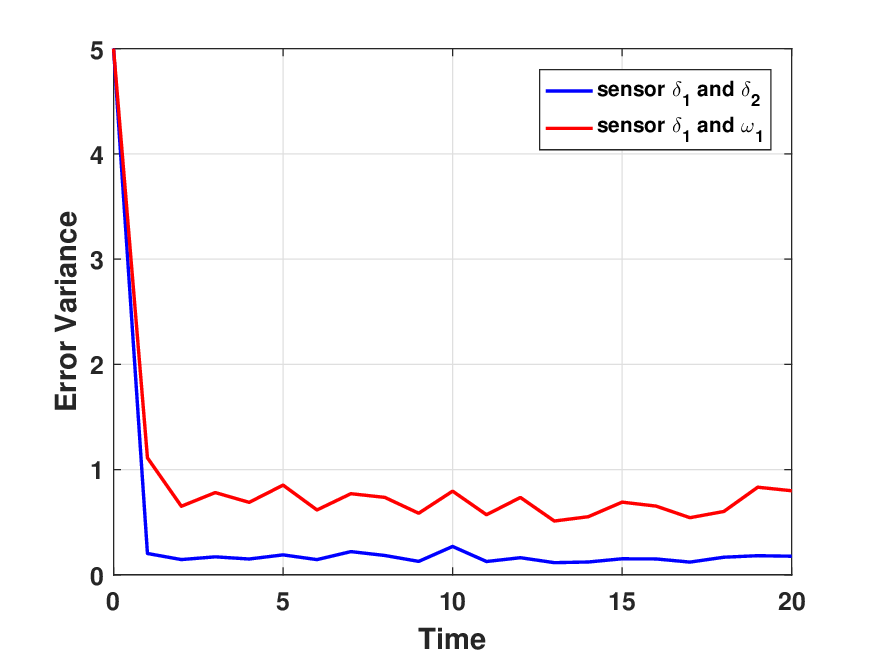}
\caption{State estimation error variance trajectories under different sensors.}
\label{sensor}
\end{center}
\end{figure}

It is easily verified
that  $(C(1), A)$ and $(C(2), A)$ are observable. However, $(C(3), A)$  is not observable and its observability matrix is
$$ W(3)= \left[ \begin{array}{cccc} 0  &  1 &         0    &     0\\
    7.7926 &  -0.1053 &  -7.7926   &      0\\
   -0.8203 &   7.8037  &  0.8203  & -7.7926\\
  219.6761 &  -1.6417 & -219.6761  &  2.2056\end{array}\right]
$$
with rank $3$.  The kernel of $W(3)$ is $M_3=[-0.7071, 0, -0.7071, 0]^{\top}$. One choice of the transformation matrix $T_3=[M_2,N_3]$
is
$$
N_3\! =\!\!\left[ \begin{array}{ccc}
     0   &  0  &   0\\
     1   &  0  &   0\\
     0   &  1  &   0\\
     0   &  0  &   1
     \end{array}\right]\!\!,\
     T_3\!=\![M_3,N_3]\!=\!\!\left[ \begin{array}{cccc}
   -0.7071    &     0    &     0   &      0\\
    0    & 1   &      0   &      0\\
   -0.7071   &      0   & 1 &        0\\
    0      &   0    &     0   & 1\end{array}\right]\!\!,
   $$
which results in
$$
  T_3^{-1} = \left[ \begin{array}{cccc}  -1.4142   &      0  &  0    &     0\\
  0 &    1 &   0 &         0\\
   -1 &        0  &  1 &        0\\
    0 &         0   &      0  &  1 \end{array}\right],$$
    and
    $$ G_3=\! \left[ \begin{array}{cccc}  \!-1.4142   &      0  &  0    &     0\!\end{array}\right],\
  F_3=\!\left[ \begin{array}{cccc}
  0 &    1 &   0 &         0\\
   -1 &        0  &  1 &        0\\
    0 &         0   &      0  &  1 \end{array}\right].
$$
We choose the desired closed-loop poles to be $\lambda=[-10.8,-8.1,-9,-9.9]$, and the packet delivery
ratio to be $\rho_1=0.99, \rho_2=0.995$. 
Fig. \ref{sensor} shows that the estimation error is affected by sensor selections.

\subsection{IEEE 33-Bus System}

We now use the IEEE 33-bus system \cite{Wu, Wong} to further illustrate our framework, methods, and algorithms in a more general setting. The original IEEE 33-bus system is a distribution network, which has been extensively used in evaluating renewable energy distribution networks, see \cite{enhanced33bus} and the references therein.

In general, the random contingencies can be any random sudden changes in system structures and parameters, such as random packet loss in communication channels in PMUs, random sensor failures, random tripping of circuit breakers, line faults, surge protection devices, etc. In this case study, we use sensor/communication interruptions for illustration. It is noted that our approach is especially useful for distribution networks with high penetration of distributed energy resources (DERS). These DERS create more dynamic buses due to local solar and wind turbines, controllable loads, etc. Such systems utilize more communication systems (cyber physical systems) in management and experience more frequent and random contingencies such as load and generation interruptions due to intermittent solar and wind systems, etc.   Our methods form an important framework to provide sustained monitoring  and estimation capability of the internal states for control and management, even in the face of such interruptions.

\textbf{1) Modeling and Linearization of the IEEE 33-Bus System}

\begin{figure*}[htb]
\setlength{\unitlength}{0.1 in}
\begin{center}
\includegraphics[height=5.5cm]{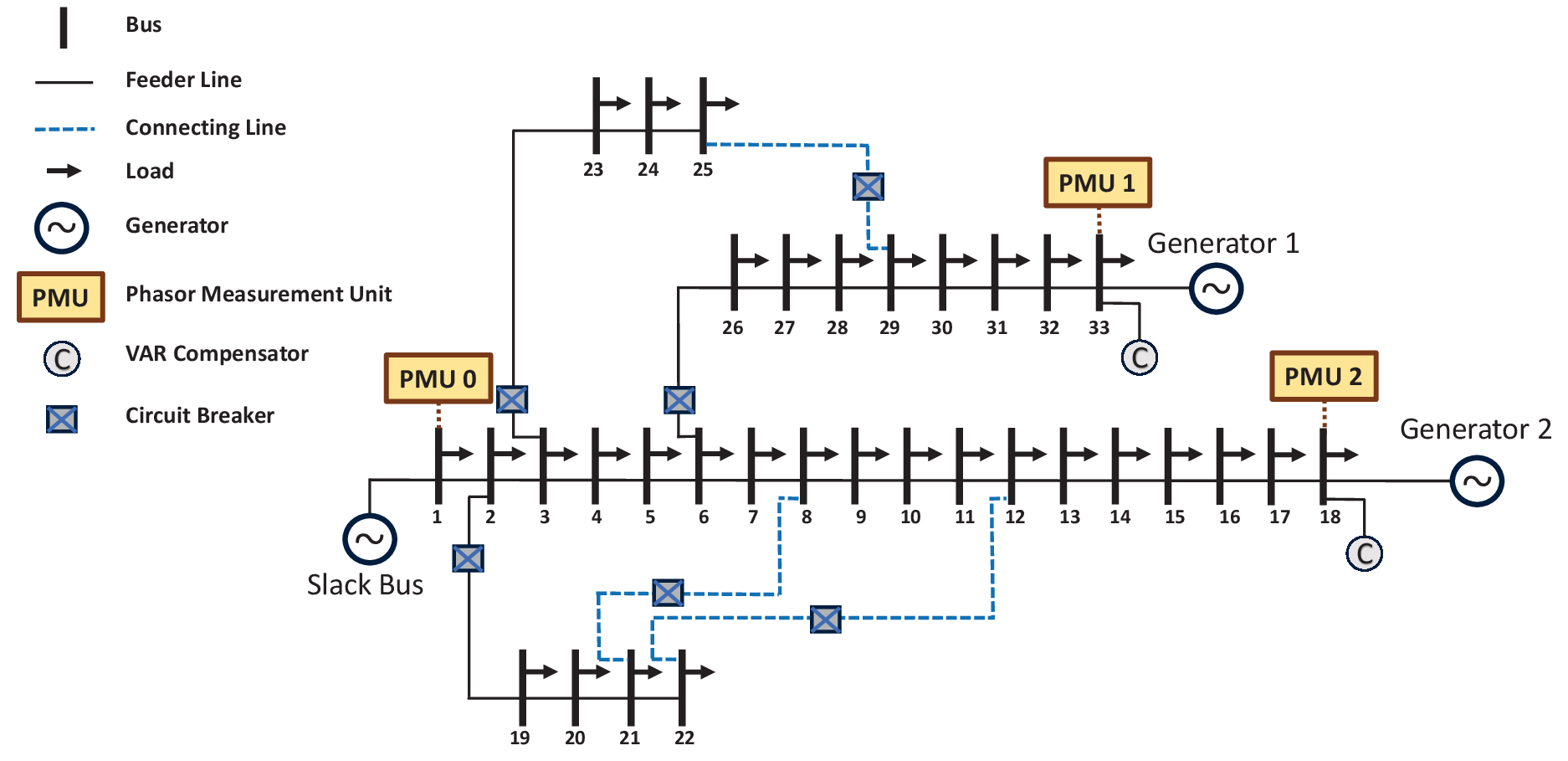}
\caption{The enhanced IEEE 33-bus distribution test system.}
\label{33bussystem}
\end{center}
\end{figure*}

The original 33-bus system \cite{Wu} contains one slack bus tied to the large grid and the rest buses are PQ-type load buses. For evaluation of renewable systems, more local generators have been added. Following the enhanced 33-bus evaluation system proposed in \cite{enhanced33bus}, in this simulation study, Bus 1 remains as a slack bus and two generators are added, at  Bus $18$ and Bus $33$,  shown in Fig. \ref{33bussystem}. The generator buses are dynamic buses whose local state space models for real power management are represented by their swing equations.  All other buses remain as PQ-type load buses as in the original configuration and non-dynamic. The slack bus voltage is set as the reference bus with constant voltage  $1\angle 0$ (pu), whose $P$ and $Q$ injections are unlimited and instantaneous in balancing powers in each step. Consequently, the slack bus is non-dynamic. All bus and load parameters are from the power flow data in \cite{Wu} and obtained from the 33-bus case file  in MATPOWER \cite{Wong, MATPOWER1, MATPOWER2}.
The base power of the IEEE 33-bus system is $100$ (MW) and the base voltage is $V_B= 230$ (kV).

The nonlinear dynamic models for Bus 18 and Bus 33 are summarized below.
Denote $\omega_{18}=\dot{\delta_{18}},\omega_{33}=\dot{\delta_{33}}$. The dynamic systems are
\bea M_{18} \dot \omega_{18} + g_{18}(\omega_{18})\ad = P^{18}_{in} - P^{18}_{out},\\
M_{33} \dot \omega_{33} + g_{33}(\omega_{33})\ad = P^{33}_{in} - P^{33}_{out},
\eea
where $P^i_{out}$ is the total transmitted power from Bus $i$ to its neighboring buses. Denote the line admittance $Y_{ij}=|Y_{ij}|\angle \gamma_{ij}$ and shunt admittance $Y_{i}=|Y_{i}|\angle \gamma_{i}$.

Since Bus 18 (and Bus 33) has only one neighboring Bus 17 (and Bus 32), we have
\begin{small}\beq{Pout}
P^{18}_{out}= V^2_{18}|Y_{18}|\cos (\gamma_{18})+ V_{18}V_{17} |Y_{18,17}|\cos (\delta_{18}-\delta_{17} - \gamma_{18,17}),\eeq
\end{small}
$\!\!$and similarly for Bus 33. The damping term is assumed to be linear with $b_{18}=0.22$, $b_{33}=0.12$. The normalized inertias are
$M_{18}=1.8$ and $M_{33}=0.9$.

The generator buses are PV buses whose power inputs $P^{18}_{in}$ and $P^{33}_{in}$ are given control inputs. As in the traditional power flow analysis, since each bus has $4$ variables ($P,Q,V,\delta$), the system contains $132$ variables. At the steady state, $P^{18}_{in} = P^{18}_L+P^{18}_{out}$ and $P^{33}_{in} = P^{33}_L+P^{33}_{out}$. From the given PV values for the generator buses, $V$ and $\delta$ for the slack bus, and PQ values for the load buses, the remaining $66$ variable can be obtained from the standard $66$ algebraic power flow equations. Under the generation powers $P^{18}_{in}=1.29$ pu and $P^{33}_{in}=0.89$ pu, the equilibrium point (the stationary operating condition) is calculated as $\bar \delta_{18}=-0.01$ (degree), $\bar \omega_{18}=0$, $\bar \delta_{33}=0.12$ (degree), $\bar \omega_{33}=0$.
The slack bus provides real power $3.94$ pu.

On the other hand, during the transient time period,   dynamic systems of the generators and algebraic network equations for power flows interact and  introduce a new iteration scheme.  During the transient time, $P^{18}_{in} \neq P^{18}_L+P^{18}_{out}$ which drives changes in $\delta_{18}$. The new $\delta_{18}$ then enters power flow analysis to result in new power flow status, including the new $P^{18}_{out}$ to the grid; similarly for Bus $33$. As a result, during transient calculation of the power flow status, there are $66$ dependent variables in power flow calculation via MATPOWER, denoted as $Z=[P^d_{out}, Q^d_{out}, P_s, Q_s, V^{nd}, \delta^{nd}]$, where the superscript $d$ refers to the dynamic buses $18$ and $33$, $nd$ refers to the non-dynamic load buses $2-17$ and $19-32$, and the subscript $s$ refers to the slack bus $1$. The rest $66$ variables are independent variables. $Z$ will be solved via $66$ algebraic power flow equations.

 The corresponding values of $Z$ at the equilibrium point are denoted by $\bar Z$.

Denote $x=(\delta_{18},\omega_{18},\delta_{33},\omega_{33})$, $u=[P^{18}_{in},P^{33}_{in}]$, the state equation is
\[ \dot x= f_0 (x, G(\delta_{18},\delta_{33}),u)=f(x, u).\]
Then the Jacobian matrix at the equilibrium point $x=\bar x,Z=\bar Z$  is
\begin{small}\begin{align*}
    &A= \frac{\partial f(x,u)}{\partial x^{\mathrm{T}}} \Bigg|_{x=\bar x}\\
    =&  \frac{\partial f_0(x,Z,u)}{\partial x^{\mathrm{T}}} \Bigg|_{x=\bar x,Z=\bar Z}\!\! + \frac{\partial f_0(x,Z,u)}{\partial Z} \Bigg|_{x=\bar x,Z=\bar Z} \frac{\partial Z}{\partial  x^{\mathrm{T}}} \Bigg|_{x=\bar x,Z=\bar Z}.
\end{align*}
\end{small}

Based on the actual expressions of $f(x, u)$, the Jacobian matrix is given by
\bea A = \left[\begin{array}{cccc} 0 & 1&0&0\\
-\frac{1}{M_{18}}
\frac{\partial P^{18}_{out}}{\partial\delta_{18}}&  -{b_{18}\over M_{18}} & -{1\over M_{18}}{\partial P^{18}_{out}\over \partial \delta_{33}}  & 0  \\
 0 &  0 & 0  &  1 \\
-{1\over M_{33}}{\partial P^{33}_{out}\over \partial \delta_{18}}&  0 & -{1\over M_{33}}{\partial P^{33}_{out}\over \partial \delta_{33}}  &  -{b_{33}\over M_{33}}
\end{array} \right].\eea

Utilizing the initial 33-bus power flow data from \cite{Wu} and \cite{Wong}\footnote{For the original data of the buses, links, generators, and loads of the 33-bus system,  please refer to the public-domain case33 file in MATPOWER, see \cite{MATPOWER1, MATPOWER2}.}, and employing MATPOWER for power flow analysis along with the computational algorithms of partial derivatives, we obtain
\[A=\left[\begin{array}{cccc}       0  &  1     &     0       &  0 \\
    -1.1280  & -0.1222  &  -0.0120      &   0\\
         0    &     0      &   0   & 1 \\
    -0.0344   &      0 &   -4.4785  & -0.1333
\end{array}\right].
\]

\textbf{2) Derivation of the RSLS for the IEEE 33-Bus System}

Suppose that  $\delta_{18}$ and $\delta_{33}$ are independently measured.
$C(1)= \left[\begin{array}{cccc} 1 & 0 & 0 & 0\\  0 & 0 & 1 & 0\end{array}\right]$ for normal operation, $C(2)=[1,0,0,0]$ for failure of Sensor 2 ($\delta_{18}$ measurement only), $C(3)=[0,0,1,0]$ for failure of Sensor 1 ($\delta_{33}$ measurement only), and $C(4)=[0,0,0,0]$ for failure on both sensors. Suppose that the packet delivery ratio for Sensor 1 is $\rho_1=0.99$ and for Sensor 2 is $\rho_2=0.995$. This data acquisition scheme can be modeled by an i.i.d. stochastic process $\alpha_k \in \clS=\{1,\ldots,4\}$ with $p_1=\rho_1 \rho_2=0.98505$, $p_2=\rho_1 (1-\rho_2)=0.00495$, $p_3=(1-\rho_1) \rho_2=0.00985$, $p_4=(1-\rho_1) (1-\rho_2)=0.00005$.

\textbf{3) Performance Evaluation: Observer Design and Convergence}

Following a design approach akin to the IEEE 5-bus systems, the pole placement method is employed to derive observer feedback gains for $\alpha_k$ values of 1, 2, and 3. In the case where $\alpha_k=4$, the observer operates in an open-loop configuration due to $C(4)$ being equal to zero.
The desired closed-loop poles are selected as $\lambda=[-3.6,-2.7,-3.3,-3]$. Then the Matlab function $L_i= place(A',C'(i), \lambda)$, $i=1,2,3$, yields the suitable feedback gains and the closed-loop error dynamics with
$A_c^i=A-L_i C(i)$, $i=1,2,3$.

\begin{figure}[htb]
\setlength{\unitlength}{0.1 in}
\begin{center}
\includegraphics[height=5.5cm]{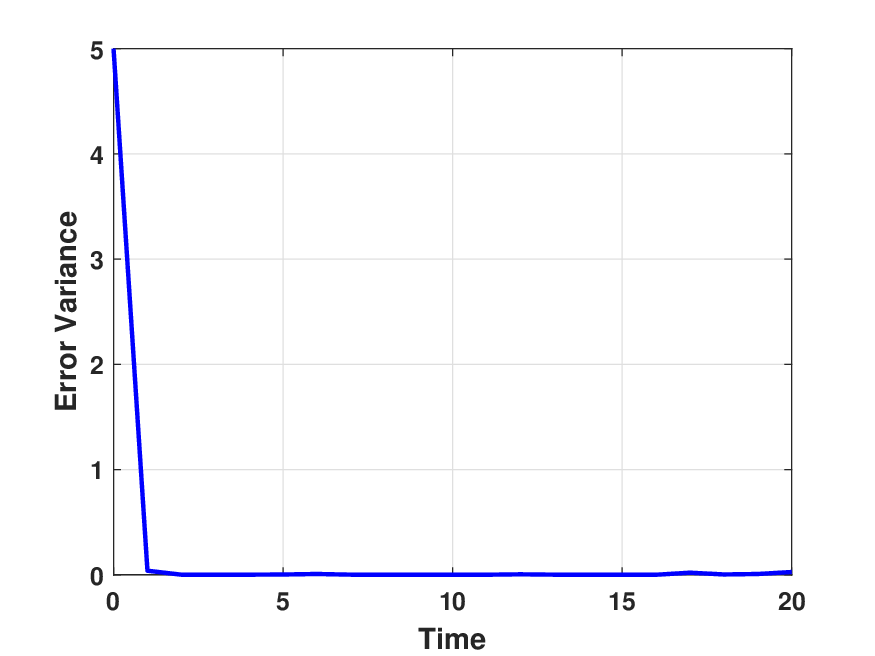}
\caption{State estimation error variance trajectories for 33-bus systems.}
\label{fig33bus}
\end{center}
\end{figure}

Select $\tau = 0.95$ (second). The initial estimation error is selected to be $e(0)=[2,0,1,0]^{\top}$ with the square of error norm $5$.
Suppose that sensor noises are i.i.d, uniformly distributed, zero mean, and the standard deviations $\sigma(1)=0.001 I_2$ on sensor configuration 1 (both sensors are functional); $\sigma(2)=0.00015$ (sensor 1 only); $\sigma(3)=0.0002$ (sensor 2 only). Estimation error (norm squared) trajectories are first recorded for 30 runs and then averaged.
Fig. \ref{fig33bus} shows the averaged estimation error trajectories on state estimation. It can be seen that
the designed observer facilitates the convergence of state estimation errors toward zero.

\color{black}

\section{Conclusions}\label{con}

This paper introduces SHS modeling for MPS. By incorporating random measurement noises and system contingencies as stochastic discrete events, we introduce a stochastic hybrid system framework for MPS that captures and utilizes dynamics of MPS to enhance the capability of internal state estimation. We present new observer design methods that combine system dynamics, stochastic information of switching processes, and noise variances to overcome the complications raised by unobservable subsystems and produce convergent observers by using only limited sensor data.  The convergence analysis and variance calculation of observation errors are established. Moreover, the trade-off between fast convergence and small steady-state error is discussed. The results of this paper indicate a promising framework of stochastic hybrid systems in dealing with sensors, state estimation, control, and management in MPS when random contingencies are taken into consideration. 

\end{document}